\documentclass{aa}
\usepackage{epsfig}
\usepackage{amsmath}
\usepackage{subcaption}
\usepackage{wasysym}
\usepackage{marvosym}
\usepackage{txfonts}
\pdfoutput=1
\usepackage{threeparttable}
\usepackage[draft]{hyperref}
\hypersetup{pdfauthor=David Doelman}
\hypersetup{backref=true, pagebackref=true, hyperindex=true, breaklinks=true,colorlinks=true,urlcolor=blue, linkcolor=blue,citecolor=blue,pagecolor=red, bookmarks=true, bookmarksopen=true}
\usepackage{epstopdf}
\hypersetup{backref=true}
\usepackage{lipsum}
\usepackage{natbib}
\usepackage{mathtools}
\usepackage{upgreek}

\bibpunct{(}{)}{;}{a}{}{,} 
\usepackage{mhchem}
\usepackage{comment}
\newcommand{\rch}[1]{\color{black}#1}

\begin{document} 

   \title{\rch{First light of a holographic aperture mask: Observation at the Keck OSIRIS Imager}}


    \author{David S. Doelman \inst{1}, Joost P. Wardenier\inst{1}, Peter Tuthill \inst{2}, Michael P. Fitzgerald \inst{3}, Jim Lyke \inst{4}, Steph Sallum \inst{5}, Barnaby Norris \inst{2}, N. Zane Warriner \inst{6}, Christoph Keller \inst{1},  Michael J. Escuti \inst{6}, Frans Snik \inst{1}} 
    
 \authorrunning{David S. Doelman et al.}
   
\institute{Leiden Observatory, Leiden University, Postbus 9513, 2300 RA Leiden, The Netherlands 
    \and 
    Sydney Institute for Astronomy, School of Physics, University of Sydney, NSW 2006, Australia
    \and
    Department of Physics and Astronomy, University of California, Los Angeles, 430 Portola Plaza, Box 951547, Los Angeles, CA 90095-1547
    \and
    W. M. Keck Observatory, 65-1120 Mamalahoa Highway,Kamuela, HI 96743
    \and 
    Department of Physics and Astronomy, University of California, Irvine, 4129 Frederick Reines Hall, Irvine, CA,
USA, 92697
    \and Department of Electrical and Computer Engineering, North Carolina State University, Raleigh, NC 27695, USA}

   \date{Received -- / Accepted --}

 
   \abstract
  {As an interferometric technique, sparse aperture masking (SAM) is capable of imaging beyond the diffraction limit of single telescopes. This makes SAM an important technique for studying processes such as planet formation at Solar System scales. However, it comes at the cost of a reduction in throughput, typically by 80-90\%.}
   {We report on the design, construction, and commissioning of a prototype aperture masking technology implemented at the Keck OH-Suppressing Infrared Integral Field Spectrograph (OSIRIS) Imager: the holographic aperture mask. Holographic aperture masking (HAM) aims at (i) increasing the throughput of SAM by selectively combining all subapertures across a telescope pupil in multiple interferograms using a phase mask, and (ii) adding low-resolution spectroscopic capabilities.}
   {Using liquid-crystal geometric phase patterns, we manufacture a HAM mask that uses an 11-hole SAM design as \rch{the} central component and a holographic component comprising 19 different subapertures. Thanks to a multilayer liquid-crystal implementation, the mask has a diffraction efficiency higher than 96\% from 1.1 to 2.5 micron. We create a pipeline that extracts monochromatic closure phases from the central component as well as multiwavelength closure phases from the holographic component. We test the performance of the HAM mask in the laboratory and on-sky. }
   {The holographic component yields 26 closure phases with spectral resolutions between R$\sim$6.5 and R$\sim$15, depending on the interferogram positions. On April 19, 2019, we observed the binary star HDS~1507 in the $\mathrm{H_{bb}}$ filter ($\lambda_0 = 1638$ nm and $\Delta \lambda = 330$ nm) and retrieved a constant separation of 120.9 $\pm 0.5$ mas for the independent wavelength bins, which is in excellent agreement with literature values. For both the laboratory measurements and the observations of unresolved reference stars, we recorded nonzero closure phases -- a potential source of systematic error that we traced to polarization leakage of the HAM optic. We propose a future upgrade that improves the performance, reducing this effect to an acceptable level.}
   {Holographic aperture masking is a simple upgrade of SAM with increased throughput and a new capability of simultaneous low-resolution spectroscopy that provides new differential observables (e.g., differential phases with wavelength).}


   \maketitle
%

\section{Introduction}

Many of the most critical aspects of stellar physics play out in a theater at Solar System scales.
These include star and planet formation, mass loss, and debris disks, to name only a few. 
High-fidelity imaging of circumstellar environments can provide key insights into these processes.
High-contrast imaging instruments with adaptive optics (AO) provide high-resolution imagery with great sensitivity, resolving many protoplanetary disks and substellar companions \citep{chilcote2018upgrading,Beuzit2019}.
However, the performance of high-contrast imaging systems is still limited by residual phase and non-common path aberrations, reducing the sensitivity \citep{macintosh2019,Beuzit2019}. 
Nonetheless, extreme AO facilities have been able to reach high contrasts ($>$14 magnitudes) down to $\sim$200 mas: a few times the diffraction limit in the near-infrared \citep{Vigan2015, nielsen2019gemini}.
For nearby populous star forming regions such as Taurus, 200 mas corresponds to $\sim 30$ AU (larger than the orbit of Jupiter or Saturn), which leaves a blind spot for critical scales of disk evolution and planet formation. 
An additional technique called sparse aperture masking (SAM), often used in concert with AO, has been able to resolve finer structures beyond the diffraction limit, for example, 20 mas at 1.65 $\mu m$ \citep{Tuthill1999}. 

Sparse aperture masking works by turning a telescope aperture into an interferometric array, in turn by using an opaque mask with small holes \citep{Haniff1987,tuthill2000michelson}.
For most applications, the holes are placed in a nondredundant fashion, which means that each baseline (the vector that connects two apertures) appears only once.
Imaging with such a mask results in an interferogram that contains many fringes with unique spatial frequencies in the image plane.
The first null of these fringes is at $0.5 \lambda/B$ instead of $1.22 \lambda/D$, where $\lambda$ is the wavelength, $B$ the longest baseline, and $D$ the telescope aperture diameter.  
A second profound advantage is the rejection of phase noise.
Non-redundancy acts to remove noise in both visibility amplitudes and phase measurements; in particular, robust observables known as closure phases have been exploited with great success. 
Closure phases are formed by taking the sum of the phases around baselines that form a closed triangle of subapertures in the pupil. 
Even before AO became well established, the robust observables delivered by SAM allowed for imaging the regions closest to stars \citep{Tuthill1999}.\\
\begin{figure*}
    \centering
    \includegraphics[width=\textwidth]{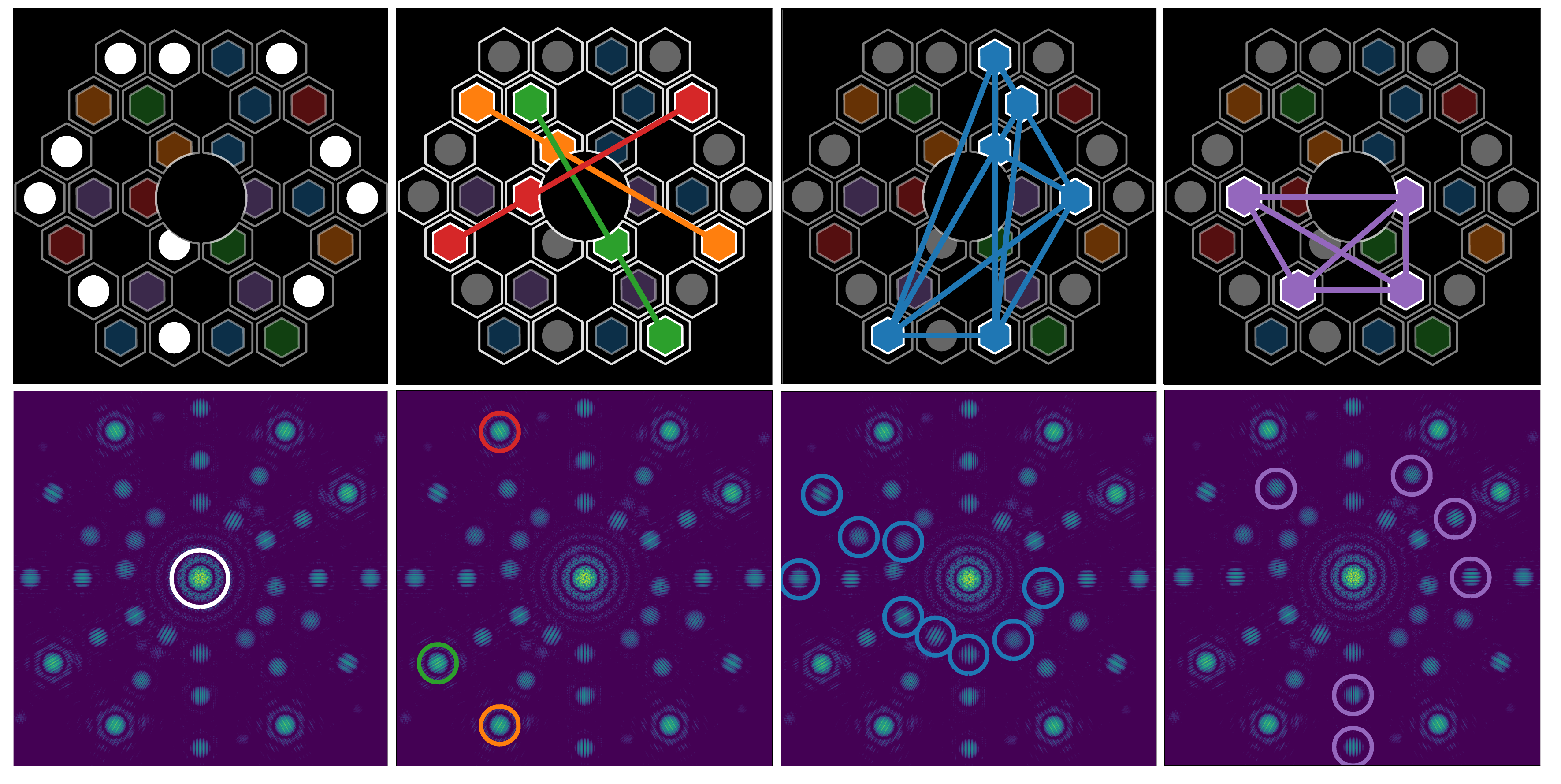}
        \caption{\rch{Design of the HAM mask for the Keck OSIRIS Imager. 
        Subapertures that are combined in one or more interferograms have the same color. 
        \textit{Top:} Subaperture combinations and baselines of the HAM mask. \textit{Bottom:} Monochromatic simulation of the HAM PSF with indications of the location of the resulting interferogram(s).
        The HAM mask incorporates a SAM mask design as the central component (leftmost panels) and adds a holographic component with off-axis interferograms (other panels).
        The baselines of the central component have been omitted for clarity.
        }}
    \label{fig:HAM_subaper}
\end{figure*}
\noindent \  Sparse aperture masking is even more powerful when used in concert with AO, providing extremely stable closure phases that result in a leap in sensitivity and contrast \citep{Tuthill2006}.
The superior calibration of closure phases in particular makes SAM more sensitive than coronagraphs for probing the smallest separations \citep[$<1-2$ $\lambda/D$; e.g.,][]{Gauchet2016,cheetham2016sparse,Samland2017}.
Furthermore, SAM has established productivity in directly resolving stellar environments, recovering dust shells \citep{Haubois2019} and structures in protoplanetary disks \citep{2008ApJ...679..762K,2019A&A...621A...7W} at Solar System scales.

The current SAM masks in Keck/NIRC2 block $80-90\%$ of the incident light.
\rch{Several different approaches have been proposed and tested to improve this throughput, for example,
the kernel phase method \citep{Martinache2010}, segment tilting interferometry \citep{monnier2009mid}, and pupil remapping interferometry \citep{Perrin2006}.
While these approaches have an increased throughput compared to SAM, they also introduce demanding system requirements, requiring high Strehl (Kernel phase), a complete overhaul in the primary mirror alignment (segment tilting), or even a completely new instrument (pupil remapping).
To the contrary, SAM is an elegant technique thanks to its simplicity: A single mask in the pupil plane adds the capability of imaging beyond the diffraction limit.}

Holographic aperture masking \citep[HAM;][]{Doelman2018} aims to increase the throughput of SAM while keeping its simplicity by using only a diffractive phase mask. 
The fundamental idea of HAM is equivalent to segment tilting \citep{monnier2009mid}, where \rch{combining} different subapertures on separate locations in the pupil allows for increasing the throughput without creating redundant baselines. 
Instead of tilted mirrors, the liquid-crystal phase mask introduces achromatic phase tilts to shift the location where subapertures are imaged onto the detector to form an interferogram.
Holographic aperture masking is implemented as an addition to a nonredundant SAM mask, with the distinction between the two components of the hybrid experiment discussed as the SAM "central component," and the off-axis HAM interferograms the "holographic component."
\rch{This is illustrated in Fig. \ref{fig:HAM_subaper}, which shows the subaperture combinations of both components for the Keck OH-Suppressing Infrared Integral Field Spectrograph (OSIRIS) Imager \citep{larkin2006osiris} and where they are imaged on the detector.
Both components provide complementary information and can be used independently of each other.
Moreover, the diffraction angle of the off-axis interferograms depends on wavelength.
When the 1D subaperture combinations are deflected to a point in the image plane orthogonal to their baseline, the wavelength smearing does not act to blur out the fringes.
Therefore, the holographic component can be designed to operate as a low-resolution spectrograph without an additional dispersing element.

 In this paper we present the first laboratory and on-sky tests of the HAM concept. We address the design of a HAM mask for the Keck OSIRIS Imager in Sect. \ref{sec:HAM_design} and the manufacturing of two HAM prototypes in Sect. \ref{sec:manufacturing}. We verify one HAM prototype in the laboratory in Sect. \ref{sec:lab_tests}, and we demonstrate the new capability of simultaneous low-resolution spectroscopy with the other prototype HAM mask with the binary system HDS 1507 in Sect. \ref{sec:on-sky}}.

\section{Design of a HAM prototype mask}
\label{sec:HAM_design}

The HAM concept shows promise for improving upon SAM mask designs, adding throughput, Fourier coverage, and wavelength diversity while keeping its simplicity. 
\rch{Here, we explore the design of a HAM phase mask for the OSIRIS Imager on Keck, shown in Fig. \ref{fig:HAM_subaper}, and how the properties of the liquid-crystal phase mask influence this design.}
\subsection{Properties of a diffractive phase mask}
\label{sec:GPH}
A critical property of the phase mask is that it needs to be able to image subapertures in off-axis interferograms.
The off-axis interferograms are rather large, with size scaling with $\lambda/D_{sub}$, where $\lambda$ is the wavelength and $D_{sub}$ is the diameter of the subaperture.
Therefore, imaging multiple interferograms onto separate locations (so as to avoid overlap) on the detector requires large phase tilts.
This makes it difficult to manufacture classical phase implementations of a HAM phase mask for transmissive pupil planes. 
A solution is offered by liquid-crystal diffractive phase masks as they have an unbounded continuous phase \citep{Escuti2016}. 
This property enables the creation of steep phase ramps that efficiently diffract light into a single order without scattering \citep{oh2008achromatic}.
In Fig. \ref{fig:HAM_subaper}, no noticeable second-order diffraction is seen for any off-axis interferogram.
Another advantage of liquid-crystal masks is that it is possible to manufacture almost any phase pattern \citep{kim2015fabrication}, meaning there is more design freedom.
We exploited this by combining phase ramps into a single phase pattern that images a single subaperture onto multiple locations in the focal plane.
This was done through multiplexing the phase ramps, and the mathematical description of multiplexing can be found in \cite{Doelman2018}.
\rch{An example of multiplexed subapertures is seen in the third column of Fig. \ref{fig:HAM_subaper}, where multiple baselines connected to a single aperture are imaged onto different interferograms.
In addition, liquid-crystal masks are diffractive because they apply a different kind of phase delay to incoming light that is independent of wavelength.
These phase delays are called "geometric phase delays" and are discussed in greater detail in Sect. \ref{sec:manufacturing}.
Due to this diffraction, the location of an imaged subaperture changes with wavelength.
The advantage of the diffractive nature is that, together with the right subaperture combination and fringe orientation, the wavelength smearing enables low-resolution spectroscopy. 
However, each interferogram can then only consist of 1D combinations of subapertures.
This limits the design freedom significantly and also greatly increases the number of off-axis interferograms. 
As shown in Fig. \ref{fig:HAM_subaper}, the fringe direction of all off-axis interferograms is orthogonal to the smearing direction for 1D combinations of subapertures.
Lastly, a specific property of liquid-crystal diffractive phase masks is that they produce two off-axis interferograms for a single phase ramp with opposite location in the focal plane. 
This can be seen in Fig. \ref{fig:HAM_subaper}, where all interferograms have an identical counterpart.
The aforementioned properties have a large impact on the design of the HAM mask, which we discuss next.}

\subsection{Phase design}

The goal of HAM is to increase the number of closure phases, throughput, simultaneous bandwidth, and spectral resolution compared to a sparse aperture mask.
\rch{A fundamental restriction of a HAM design is the size of the detector. 
The OSIRIS detector, a Teledyne Hawaii-2RG HgCdTe detector with a size of 2048$\times$2048 pixels, provides a unique opportunity.
With a plate scale of 10 mas/pixel, the detector is large enough to add many baselines to the holographic component. 
We placed the subapertures in alignment with the Keck primary mirror segmentation, such that every subaperture is centered on one segment.
The segments that are crossed by a spider are not used in the design.
Moreover, the subapertures of the holographic component are hexagonal to increase their throughput, and we increased their diameter if the smallest baselines were larger than the distance between neighboring subapertures.
The subapertures of the central component are circular as this keeps the central component point spread function (PSF) circularly symmetric, and the outer Airy rings have equal strength for all holograms at equal radius. 
Next, we briefly discuss some considerations of the presented design.}

\subsubsection{The central component}

The design of a HAM mask starts with optimizing a SAM mask as the central component.
\cite{Tuthill2018} showed that adding redundancy allows for a boost in the S/N with respect to nonredundant masking by $\gtrsim$50\%. 
\rch{Therefore, the central component of the Keck OSIRIS HAM mask is optimized to maximize throughput, the number of baselines, and the number of closure triangles, at the cost of a few redundant baselines.
The optimal choice is an 11-hole mask, as shown in Fig. \ref{fig:HAM_subaper}.
This 11-hole mask has 50 unique baselines (out of 55), which means that discarding these redundant baselines still increases the number of nonredundant baselines compared to a nonredundant nine-hole mask with 36 unique baselines. }

\subsubsection{The holographic component}
\rch{With the central component defined, 19 of the 30 available segments (63\%) are not in use. 
Combined with the possibility to multiplex each subaperture, there is an extremely large number of possible combinations.
For this prototype, we combined two different approaches, based on the balance between the flux in each interferogram and the number of baselines and closure phases. 
The first approach combines subapertures that are in a straight line, and it is limited to a maximum of three nonredundant subapertures for the Keck aperture. 
Their interferogram has a relatively high flux but yields only a single closure phase. 
We used nine of the 19 subapertures to create three of these combinations, as shown in the second column of Fig. \ref{fig:HAM_subaper}. 
The other approach combines more subapertures in multiple interferograms to make use of the rapid growth in the number of baselines and closure phases as a function of the number of interfered subapertures. 
This approach uses multiplexing to combine all subapertures with all others. 
As the number of holograms roughly increases with the number of baselines, it is critical that in this approach as well we maximize the number of subapertures that are imaged into a single interferogram. 
For example, the third column of Fig. \ref{fig:HAM_subaper} shows six subapertures but only nine interferograms, instead of 15. 
In this case we combined three subapertures into a single interferogram three times.
The subapertures are multiplexed three or four times. 
The flux in all interferograms is reduced due to multiplexing, resulting in a lower intensity compared to the other approach, a reduction of a factor of nine to 12. 
However, the number of closure phases of the ten subapertures that are combined this way is 13.
Overall, the holographic components consist of 18 holographic interferograms per polarization, yielding 16 closure triangles.
  
The subaperture locations and their mapping to the focal plane can be found in Appendix \ref{app:HAM_OSIRIS_specs}.
The full overview of HAM properties is provided in Table \ref{tab:sam_ham_overview}.
\begin{figure}
    \centering
    \includegraphics[trim=0 5 0 5,clip,width=0.5\textwidth]{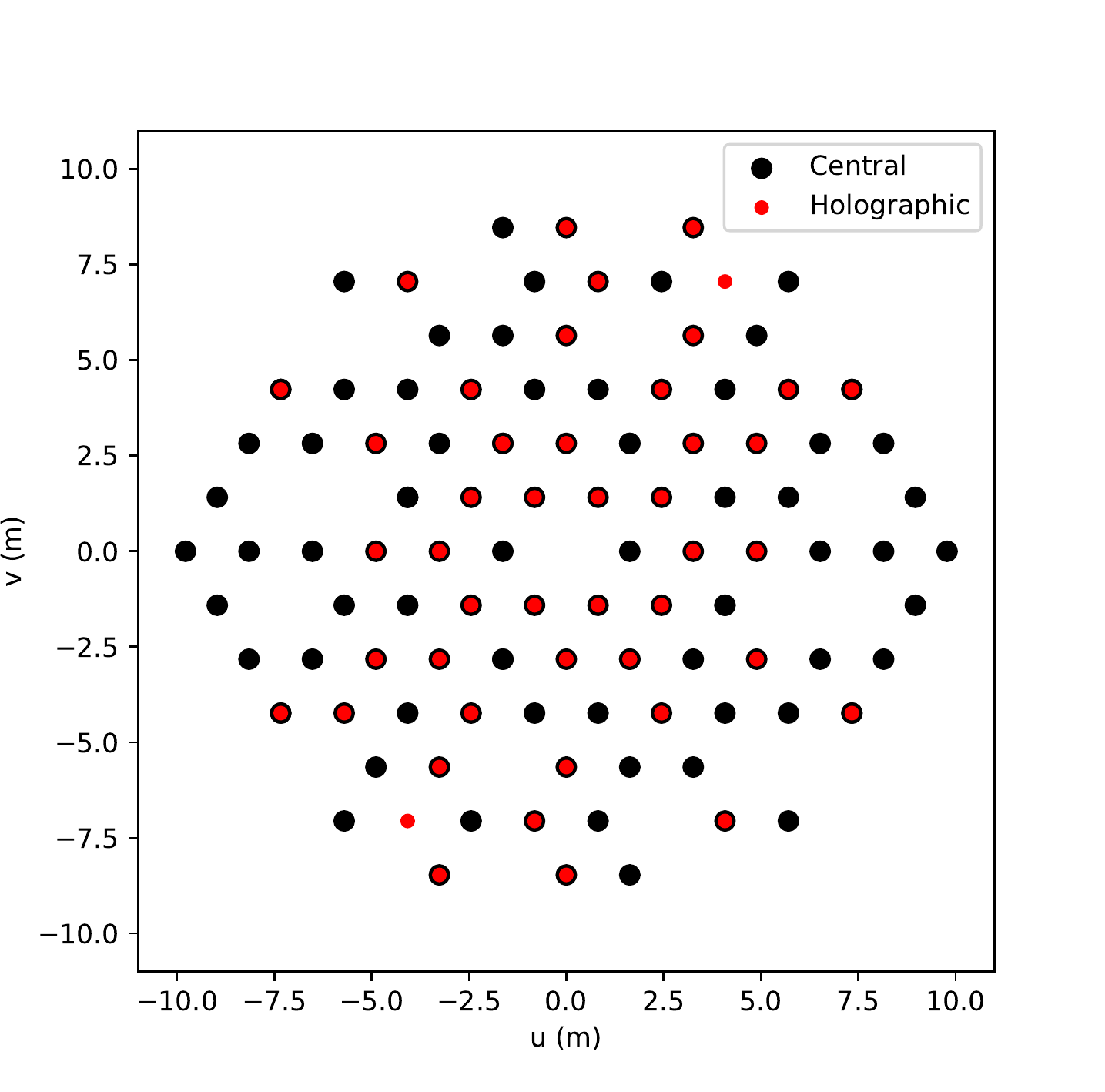}
    \caption{$uv$-coverage of both the central and holographic components.}
    \label{fig:HAM_PSF}
\end{figure}
\rch{The table} also makes a distinction between the individual contributions  from the central component and holographic component.
Several numbers are worth pointing out. 
Apart from transmission and diffraction efficiency, the total throughput of the mask is $>30\%$, which is a factor of three higher than the throughput of the central component.
Whereas the central component features ten repeated baselines (i.e., five pairs of "redundant" baselines), 13 out of the 30 HAM baselines have at least one identical counterpart. 
However, because the light is mapped onto different spots in the focal plane, their complex visibility can be computed independently. 
\begin{table}
        \vspace{-10pt}
        \centering 
        \captionsetup{justification=centering}
        \caption{Specifications of the HAM mask.}
        \vspace{-5pt}
        \begin{threeparttable}
        \begin{tabular}{@{}l c@{} c@{} c@{} }
                \hline \hline
                  & \rch{CC} \ & \rch{HC} \ & \rch{HAM} \ \\  \hline
                Subapertures & 11 & 19 & 30 \\
                Throughput & 10.1\% & 21.9\% & 32.0\% \\
                Baselines & 55 & 30 & 85 \\ 
                Unique baselines & 50 & 23 & 51 \\
                Single baselines & 45 & 17 & 27 \\
                \rch{Repeated baselines}& 10 & 13 & 58 \\
                \ \ $\bullet$ Pairs & 5 & 5 & 16 \\ 
                \ \ $\bullet$ Triplets & 0 & 1 & 7 \\
                \ \ $\bullet$ Quadruplets & 0 & 0 & 0 \\
                \ \ $\bullet$ Quintuplets & 0 & 0 & 1 \\
                \rch{Closure Triangles} &  &  &  \\ 
                \ \ $\bullet$ w/ repeated SAM baselines & 165 & 26 & 191 \\ 
                \ \ $\bullet$ w/o repeated SAM baselines & 88 & 26 & 114 \\
                \rch{Unique Closure Triangles} &  &  &  \\ 
                \ \ $\bullet$ w/ repeated SAM baselines & 165 & 26 & 190 \\ 
                \ \ $\bullet$ w/o repeated SAM baselines & 88 & 26 & 114 \\
                \rch{$uv$-points} & & & \\ 
                \ \ $\bullet$ w/ repeated SAM baselines & 100 & 46 & 102 \\ 
                \ \ $\bullet$ w/o repeated SAM baselines & 90 & 46 & 98 \\ \hline 
        \end{tabular}
         \begin{tablenotes}
      \small
      \item The mask (HAM) specifications are decomposed into contributions from the central component (CC) and the holographic component (HC).
    \end{tablenotes}
    \end{threeparttable}
        \label{tab:sam_ham_overview}
\end{table}
The HAM mask has a total of 85 baselines, 51 of which are unique. 
The unique baselines can be divided into two groups: (i) "single"} baselines, which occur only once, and (ii) repeated baselines, which occur either two, three, or five times (see Table \ref{tab:sam_ham_overview}). 
With the 50 unique baselines accommodated by the central component, the $uv$-plane is uniformly sampled and provides great coverage.
The 23 unique baselines of the holographic component provide sufficient $uv$-coverage that allows it to be used on its own.
Moreover, the baselines present in both components can be used to improve the calibration of the data.

\subsubsection{Focal plane design}
\label{subsec:focalplanedesign}
Here we switch to the focal plane to explore the effects of hologram placement in more detail.
If the subsets of apertures are chosen, the focal plane has a set of holograms that need to be given a location.
Each hologram can be imaged onto a line, where the separation with respect to the central component is a design freedom.
This impacts spectral resolution, spectral bandwidth, and interference between higher-order terms.
We define the spectral resolution of the holographic component as the maximum of two terms. 
Both terms are independent and are defined in different planes (i.e., the focal plane and the $uv$-plane) and are derived in Appendix \ref{app:spectral_resolution}.
The first term relates to the diffraction by the gratings of the HAM phase mask. 
For two subapertures with a subaperture diameter $D_{sub}$ and a phase grating with period P, the spectral resolution in the focal plane is given by
\begin{equation}
R_{fp} = \frac{\lambda}{\Delta \lambda} = \frac{D_{sub}}{1.22 P}. 
\label{eq:spec_res1}
\end{equation}
The second term is a fundamental property of interferometry, where the $uv$-points, or baselines, are wavelength-dependent due to diffraction.
If the two subapertures form a baseline, $\mathbf{b}$, their spectral resolution in the $uv$-plane is given by
\begin{equation}
R_{uv} = \frac{\lambda}{\Delta \lambda} = \frac{|\mathbf{b}|}{D_{sub}}.
\label{eq:spec_res2}
\end{equation}
With different methods we are able to fit fringes directly in the focal plane or retrieve complex visibilities from the $uv$-plane independently.
We can choose which method to apply to obtain the highest spectral resolution.
For most holograms, $\frac{D_{sub}}{1.22 P} > \frac{|\mathbf{b}|}{D_{sub}}$, yet it can be worthwhile to fill the image plane closest to the central component with the longest baselines. 
A maximum spectral resolution is obtained when the holographic interferograms are placed near the edges of the detector. 
\rch{For this reason, we put the linear non-multiplexed combinations at the largest separation with the central component.}
However, doing so introduces a larger sensitivity to the effects of \rch{imperfect} optics.
This is particularly problematic for closure triangles where the baselines are imaged at different parts of the detector. 
For example, image distortion can change the shape and location of the holographic interferograms.
Variable image quality across the focal plane can add different phase offsets to each baseline, resulting in a nonzero closure phase.
A full analysis of these effects is beyond the scope of this paper.
\begin{figure*}
    \centering
    \includegraphics[width=\textwidth]{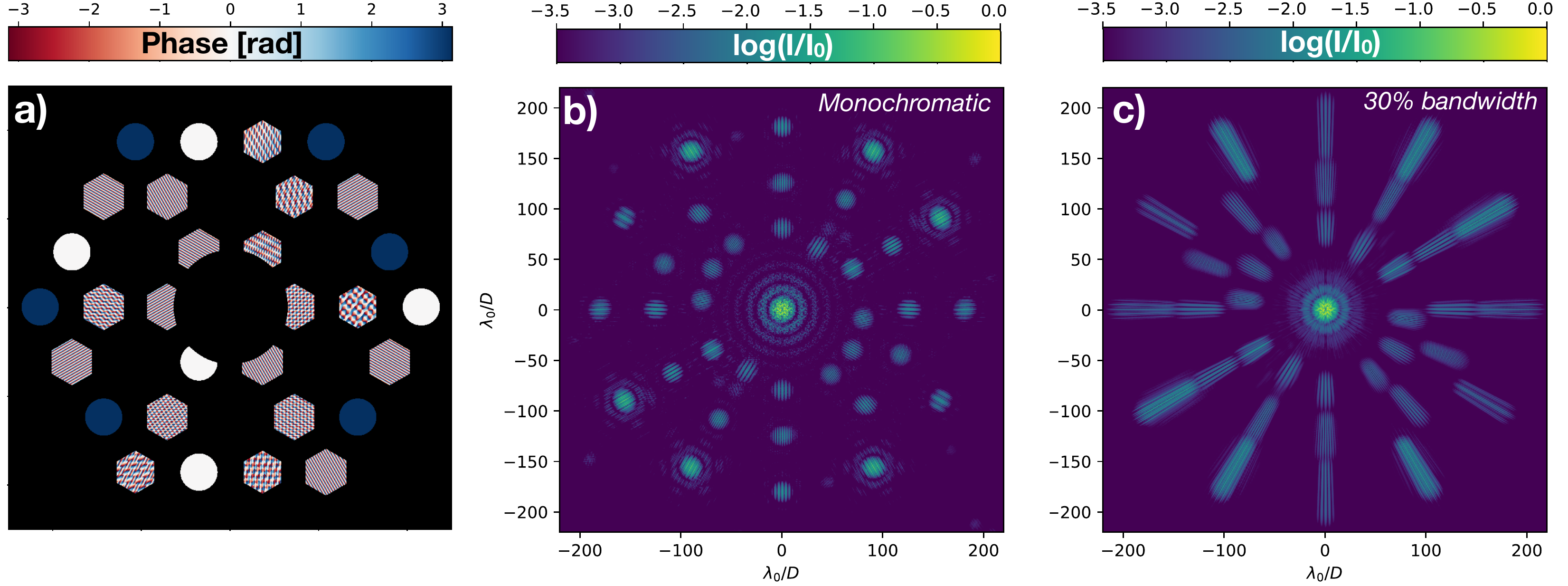}
    \caption{Design of the HAM mask for OSIRIS. $(a)$ Phase pattern of the HAM optic, masked by the amplitude mask. $(b)$ Simulated monochromatic PSF. $(c)$ Simulated PSF with 30\% bandwidth.}
    \label{fig:HAM_design}
\end{figure*}
\subsubsection{Dynamic range}

The central component is much brighter than the holographic component as all subapertures of the central component constructively interfere on the optical axis.
\rch{We reduced the peak flux in the central component by nulling the peaks with $0$ or $\pm\pi$ phase offsets. 
This reduces the dynamic range of a single image. 
The central component has six subapertures with a $\pi$ phase offset and five with a $0$ phase offset. 
The introduced phase offsets are achromatic and therefore achieved over the full wavelength range of the OSIRIS instrument. 
Moreover, these phase offsets change some closure phases by introducing a static offset.
This is taken into account in the data reduction pipeline.
We note that this specific combination of offsets was selected to provide nulling in the PSF without nulling doubly redundant baselines. 
The final phase pattern and PSF are shown in Fig. \ref{fig:HAM_design}.
The holograms do not overlap for bandwidths below 30\%.} \\

\section{Manufacturing of two prototypes}
\label{sec:manufacturing}

\rch{The diffractive HAM masks are manufactured as patterned liquid-crystal optics \citep{Escuti2016}.
Other examples of patterned liquid-crystal optics are the vector vortex coronagraph \citep{mawet2009optical} and the vector-apodizing phase plate coronagraph \citep{Snik2012}.
These liquid-crystal optics are half-wave retarders with varying fast-axis orientation.
When circularly polarized light travels through such an optic, it acquires geometric phase (or Pancharatnam-Berry phase), which is different from the classical phase that arises from optical path differences.
The geometric phase, $\phi (x,y)$, only depends on the geometry of the fast-axis orientation, $\theta(x,y)$, that is,\begin{equation}
\phi(x,y) = \pm 2 \theta(x,y),
\label{eq:geometric-phase}
\end{equation}
where the sign is determined by the handedness of the circular polarization state of the incoming light and $x,y$ indicates the pupil-plane coordinates \citep{Escuti2016}.
The fast-axis orientation for a phase ramp is shown in Fig. \ref{fig:image_optic}.
The geometric phase is independent of wavelength, as assumed in Sect. \ref{sec:HAM_design}.
Unpolarized light is defined as having no preferred state of polarization and contains, on average, equal amounts of left- and right-circular polarization.
Therefore, two off-axis holograms are created for unpolarized light going through a phase ramp of a subaperture because the phase has an opposite sign for both circular polarization states.  
Next we summarize how these liquid-crystal optics are manufactured.}

\subsection{Manufacturing of liquid-crystal optics and the emergence of polarization leakage}
\label{sec:leakage_influence}
Manufacturing a diffractive phase mask requires control of the fast-axis orientation and requires the tuning of the retardance to be half-wave.
Both of these properties can be controlled to a very high degree with liquid-crystal technology.
With a direct-write method, almost arbitrary phase patterns can be written in a photo-alignment layer \citep[PAL;][]{Miskiewicz2014a}. 
Birefringent liquid-crystal layers deposited on the PAL keep this orientation pattern due to spontaneous self-alignment. 
Changing the retardance is possible by stacking these \rch{layers, each with an optimized \rch{thickness} and twist,} into a monolithic film \citep{komanduri2013multi,kim2015fabrication}.
By tuning these parameters, these "multi-twist retarders" are capable of achieving high diffraction efficiencies over large bandwidths.
The layers are cured with UV radiation, and the liquid-crystal film therefore constitutes a static phase pattern. 
Moreover, the optic is completely flat and can easily be combined with an amplitude mask.

\rch{Writing inaccuracies of the orientation pattern lead to changes in the phase pattern.
A deviation from half-wave retardance has a more severe impact on the performance of HAM. 
The fraction of light that acquires geometric phase depends on the retardance, where half-wave retardance yields close to 100\% diffraction efficiency. 
The fraction of light that does not acquire geometric phase is called polarization leakage and is, apart from a global piston term, unaffected by the optic.
It is highly redundant as all subapertures of the HAM mask are represented in the polarization leakage term.
This term is imaged onto the location of the central component and can severely increase the noise on closure phase retrieval. 
This noise is dependent on the polarization state of the incoming light.
For unpolarized light or circularly polarized light, the visibilities of the central component are now the sum of the visibilities from the SAM PSF and the leakage PSF, that is, 
\begin{equation}
V(\mathbf{f}) = c_V^2 V_{SAM}(\mathbf{f}) + c_L^2 V_{leak}(\mathbf{f}),
\label{eq:visibilitysum}
\end{equation}
where $c_V$ and $c_L$ are scaling factors that only depend on the retardance of the HAM optic. 
For small offsets in the retardance, $c_L^2 \ll c_V^2$, yet the impact can still be significant as $V_{leak}(\mathbf{f})$ is influenced by the many redundant baselines.
Phase aberrations impact $V_{leak}(\mathbf{f})$; as such, the central component is sensitive to these aberrations and so are the closure phases.
For linear polarization or instrumental crosstalk, $V(\mathbf{f})$ also contains cross-terms from interference between the leakage PSF and central component PSF.
The impact of the polarization state and phase aberration on the HAM OSIRIS design is simulated in Appendix \ref{app:leakage}, and a solution is presented in Sect \ref{sec:conclusions}.}

\subsection{Prototypes HAM v1 and HAM v1.5}
\begin{figure}
    \centering
    \includegraphics[width=0.5\textwidth]{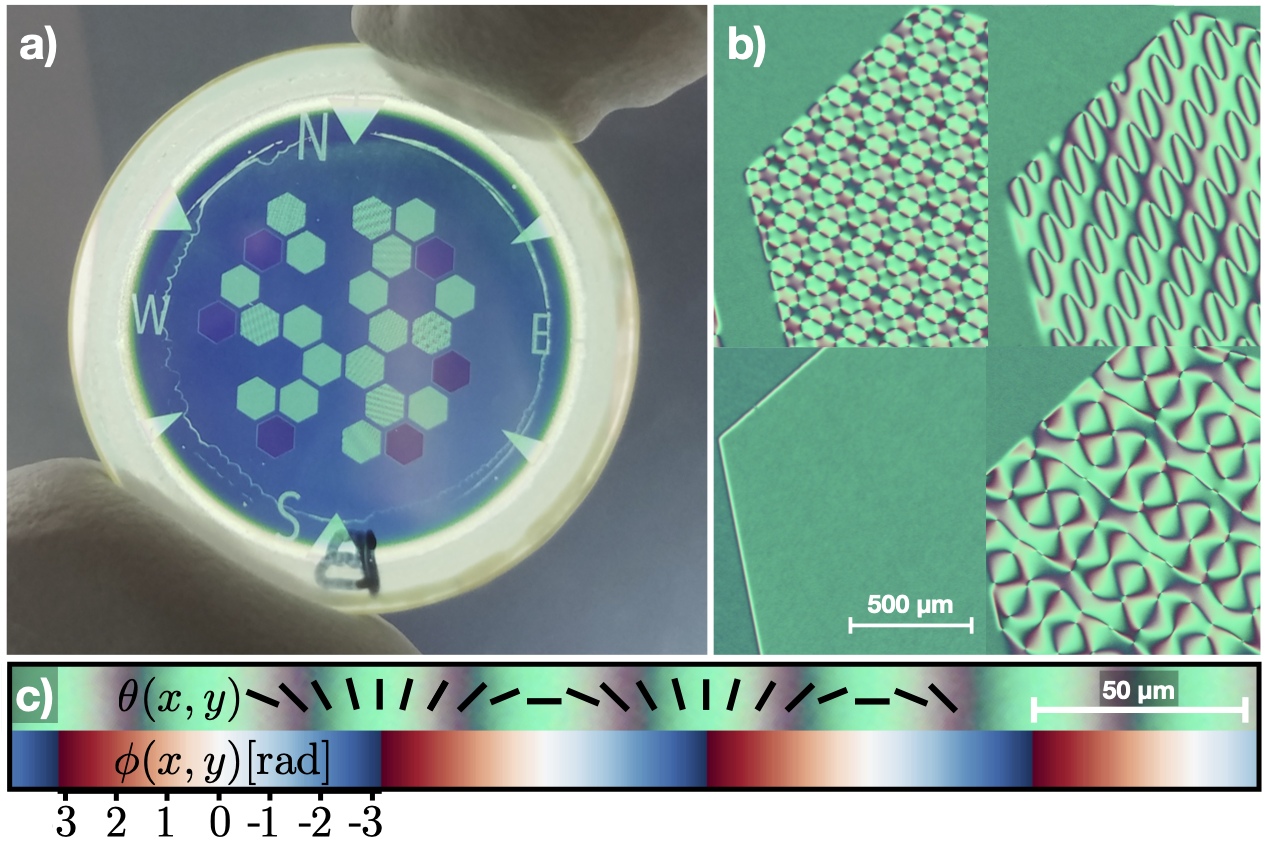}
    \caption{Images of the HAM optic between polarizers at different scales. \rch{The black lines in panel (c) indicate the local fast-axis orientation, $\theta (x,y)$, assuming parallel polarizers. The corresponding geometric phase, $\phi (x,y)$, is indicated below for one polarization state.}  Image credit: ImagineOptix. }
    \label{fig:image_optic}
\end{figure}

Two prototype HAM devices were manufactured by ImagineOptix in August 2018, labeled part A and part B. 
The first HAM device, part A, was manufactured using a 1 inch flat \rch{calcium fluoride (CaF2)} substrate with a thickness of 5 mm, while the second HAM device, part B, was fabricated using a 1 inch wedged CaF2 substrate with a thickness of 1 mm. 
The front sides have the same three-layered liquid-crystal multi-twist retarder film, aimed at minimizing polarization leakage between $1$ and $2.5$ $\mu$m.
Both devices have an antireflection coating for this bandpass on the backside of the substrates.
The phase pattern with a diameter of 25.4 mm was generated with 5 micron pixels.
The polarization leakage measured by the manufacturer is less than 3\% between $1$ and $2.5$ $\mu$m.
Additional alignment markings have been added to the pattern, outside of the pupil diameter of 13.5 mm. 
An image of the optic between polarizers is shown in Fig. \ref{fig:image_optic}a. 
In addition, the optic was inspected under a microscope between polarizers. 
The four microscope images presented in Fig. \ref{fig:image_optic}b show the high quality of the manufacturing process. 
Amplitude masks were laser-cut in 100 $\mu$m brass and/or 304 stainless steel foils with a diameter of 20.83 mm using a picosecond laser machining facility (OptoFab node of ANFF, Macquarie University, Sydney).
It was screwed in place in a holder in the filter wheel assembly. 
\noindent The first version (HAM v1) was installed in the imaging arm of OSIRIS at the Keck I telescope in September 2018. Each position of its first filter wheel contains separate pupil mask and filter mount assemblies. 
The amplitude mask was installed in the pupil assembly facing the incoming beam, while the HAM optic was installed in the opposite 1 inch filter side of the wheel assembly. 
Therefore, a gap of several millimeters was present between the amplitude mask and the HAM optic.
This version was tested with an internal source in OSIRIS in April 2019. 
The results are presented in Sect. \ref{sec:on-sky}.

\begin{figure}
    \centering
    \includegraphics[width=0.45\textwidth]{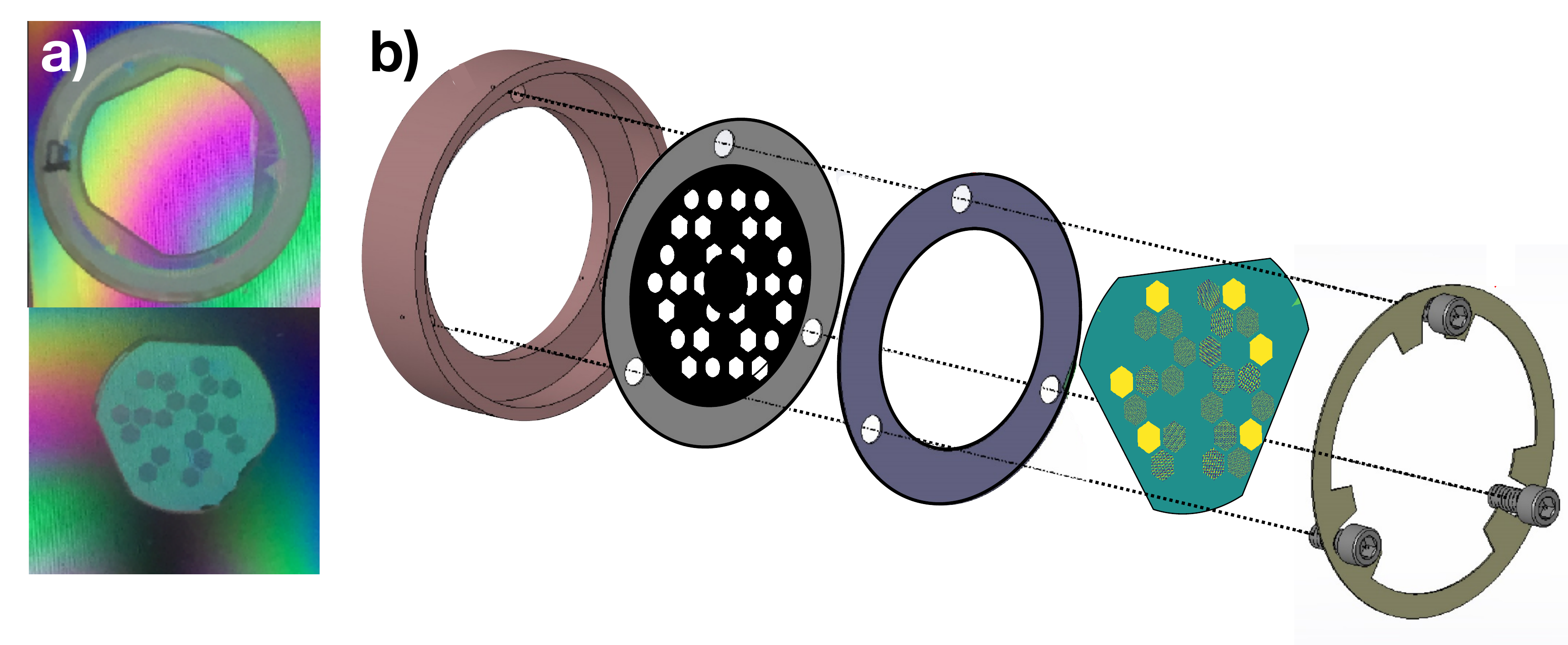}
    \caption{Manufacturing of HAM v1.5. $(a)$ Images of the diced HAM v1.5 phase optic. $(b)$ Assembly of HAM v1.5 in the OSIRIS pupil mount.}
    \label{fig:HAM_assembly}
\end{figure}

An updated version (HAM v1.5) was created to allow the HAM optic and the amplitude mask to be installed in the same pupil mask holder, reducing the separation between the optics to almost zero. 
The HAM optic of HAM v1.5 is a cutout version of the spare HAM v1 phase mask (part B). 
The cutout mask and assembly of the optic in the mount are shown in Fig. \ref{fig:HAM_assembly}. 
Initial laboratory tests were conducted in Sydney in \rch{October} 2019. HAM v1.5 was installed in OSIRIS in February 2020, replacing HAM v1. 
We present the results of the laboratory tests in the next section.
\section{Laboratory tests of HAM v1.5}
\begin{figure}
    \centering
    \includegraphics[width=0.45\textwidth]{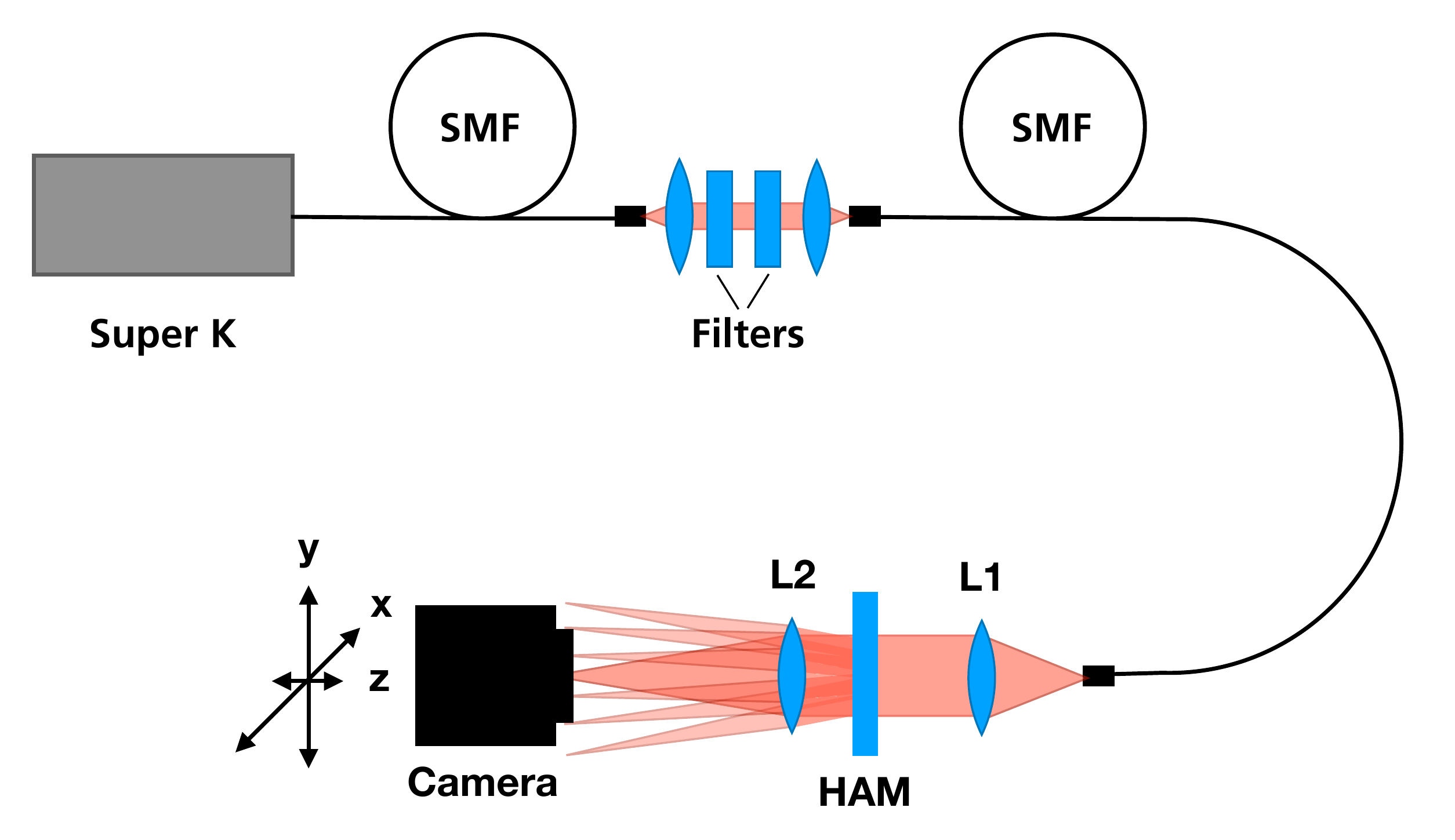}
    \caption{Laboratory setup used to characterize the HAM v1.5 optic.}
    \label{fig:HAM_lab_setup}
\end{figure}
\label{sec:lab_tests}
\begin{figure*}
    \centering
    \includegraphics[width=\textwidth]{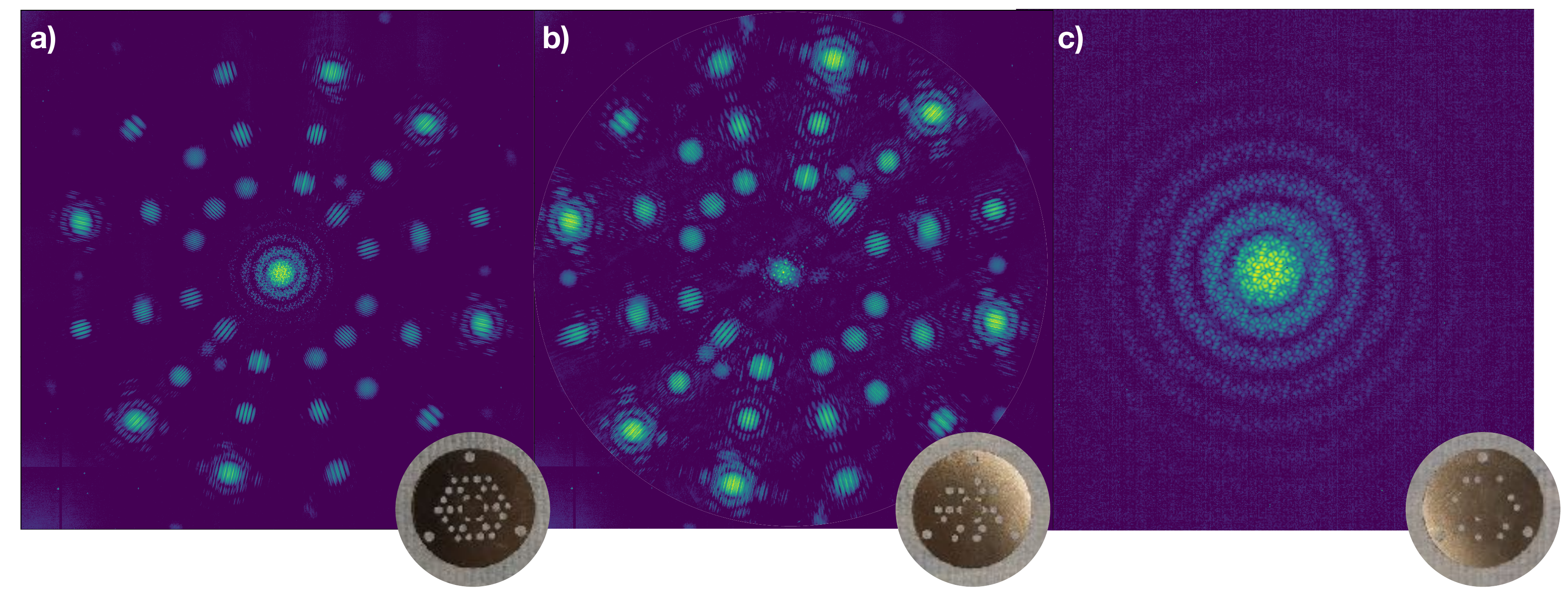}
    \caption{Images of the normalized HAM PSF with different amplitude masks at 1400 nm. $(a)$ Full HAM PSF showing that all holograms are present and contain interferograms. $(b)$ HAM PSF of the holographic component where the central term is a zero-order diffraction term of the holographic gratings. $(c)$ Close-up of the central component PSF.}
    \label{fig:HAM_PSF_lab}
\end{figure*}
We tested the HAM v1.5 optic in the laboratory using the setup described in Fig.~\ref{fig:HAM_lab_setup}.  
The light source is a SuperK COMPACT from NKT Photonics and is connected to a custom reimaging system through a single-mode fiber (SMF). 
The reimaging system allows us to insert spectral filters and neutral density filters in a collimated beam before injection into the optical setup with a second SMF.
We used filters from the Thorlabs IR Bandpass Filter Kit from 1000 nm to 1600 nm. 
Light from the second SMF is collimated with a Thorlabs 1 inch doublet with a focal length of 150 mm (AC254-150-C-ML). 
A second 300 mm (AC254-300-C-ML) doublet is placed close to the HAM optic to prevent the vignetting of individual HAM apertures.
The camera is a CRED2 and is mounted on motorized X,Y,Z translation stages to fully capture the HAM PSF. 
We used this motorized camera because a 2K pixel science grade detector, such as a Teledyne Hawaii-2RG HgCdTe detector, did not fit in the budget of the laboratory experiment.
The full PSF is captured by recording images on a 5x5 grid. 
With this sampling, neighboring images now overlap significantly, and this overlap is used for a more precise image registration between images.
Both the CRED2 and the translation stages are controlled by a Matlab script that can capture images of the PSF. 
The background is estimated from the median of 100 images \rch{in which} the source is turned off.
The background is captured before the image sequence, where for every position ten images are averaged, and the background is subtracted.
\subsection{The point spread function}
The translation between neighboring images are extracted using the scikit-learn \texttt{feature.register\_translation} function on a masked PSF.
Individual images are shifted accordingly and stored in a 3D array, where the first axis corresponds to the number of images and the other two to the x and y positions in the combined image.
The final mosaic is the median along the first axis, which can be the single pixel value if a PSF region is imaged only once, or the median of multiple values when there is overlap.
We remark that this method relies heavily on PSF stability, especially as fringes are the features used to align images with respect to each other. 
Any change to the fringe phase could lead to minor misalignments of images with respect to each other. 
We assume this effect is small as the overlapping region between neighboring images contains multiple interferograms with different orientations.
Any shift of the fringe phase in a single spot would not throw off the alignment.
A passive setup can only generate changes in many closure phases simultaneously when optics move, which does not occur on timescales of two consecutive images.
However, it demonstrates the sensitivity of fringe phases to the alignment of images with respect to each other. 
In addition, the PSFs are Nyquist-sampled for the shortest wavelengths, meaning that even small sub-pixel shifts lead to large phase offsets. 
This is only a limitation of the laboratory setup and will not affect the performance of HAM in the OSIRIS instrument, where the PSF is fully captured by the camera.

\begin{figure}
    \centering
    \includegraphics[width=0.45\textwidth]{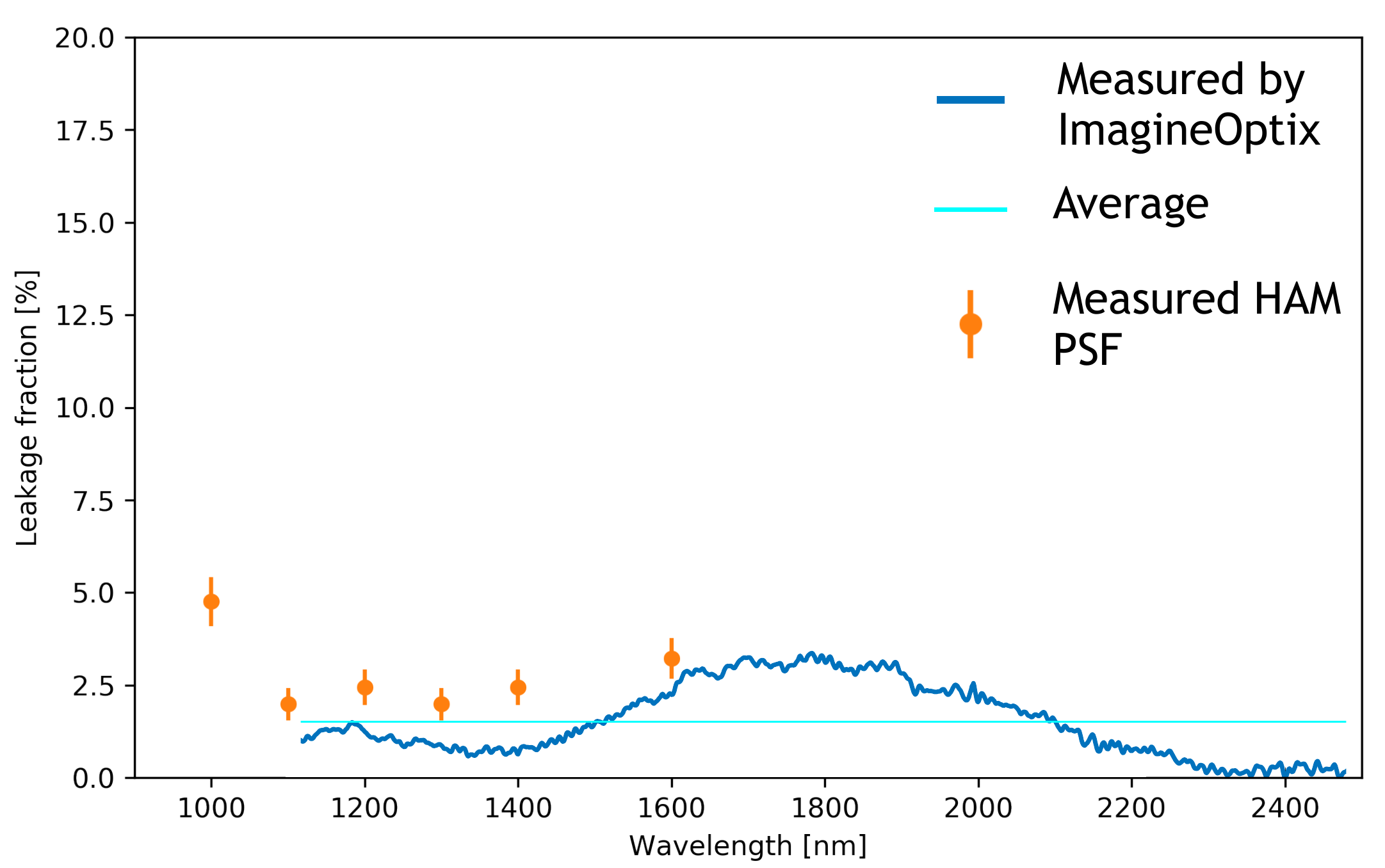}
    \caption{Measured zero-order leakage fraction of the HAM v1.5 optic.}
    \label{fig:HAM_leakage}
\end{figure}

We recorded images of the HAM PSF using three different amplitude masks.
The different amplitude masks are the full HAM amplitude mask and two masks to isolate only the central and holographic components of the PSF.
This allows us to analyze their individual PSFs and characterize the zero-order leakage of the HAM phase optic. 
The three masks and their corresponding PSFs at 1400 nm are presented in Fig. \ref{fig:HAM_PSF_lab}.
The HAM PSFs closely resemble the simulated PSFs, and they confirm that the HAM optic was not damaged during the dicing process. 
The holographic component directly shows the zero-order leakage. 
Using \texttt{HCIPy}, we forward-modeled the HAM PSF with variable retardance.
Changing the retardance changes the intensity of the central leakage term with respect to the holographic interferograms.
By minimizing the normalized difference of the modeled and measured PSFs, we \rch{we can extract} the retardance at different wavelengths. 
The fitting results are shown in Fig. \ref{fig:HAM_leakage}, together with measurements of the efficiency of a polarization grating with the same liquid-crystal recipe. 
There is an offset of roughly 1\% between these two methods of measuring the zero-order leakage. 
A reduced diffraction efficiency for multiplexed gratings as compared to single gratings could explain this difference in zero-order leakage intensity.
Multiplexing gratings with high frequencies leads to local phase patterns with extreme phase gradients, which might not be fully captured by the direct-write method.
If that is the case, the diffraction efficiency is reduced.
However, the presented measurements are unable to distinguish a change in diffraction efficiency due to multiplexing from a change in retardance.
A consequence of the increased zero-order leakage is an increased sensitivity of the closure phases to polarized light, as discussed in Sect. \ref{sec:leakage_influence}.
\subsection{The data reduction pipeline}
\label{sec:pipeline}
\rch{Here, we briefly discuss the new pipeline that was developed for HAM data reduction. 
The HAM pipeline is written in Python using the \texttt{HCIPy} package \citep{Por2018}. 
All operations are carried out separately for the central and holographic components.
The phase and amplitude of the complex visibility are measured on each baseline. The pipeline implements both a fringe-fitting method \citep[e.g.,][]{Lacour2011a, Greenbaum2014} and a Fourier method  \citep[e.g.,][]{Tuthill1999, Sallum2017} to retrieve the complex visibilities.
For the fringe-fitting method, we apodized the PSF of the central component using a power-two super-Gaussian window function with a full width at half maximum (FWHM) of 110 $\lambda/D$ in both axes. 
This separates the central component from the holographic component, in addition to suppressing high-frequency noise in the $uv$-plane.
Similarly, for the holographic component, we apodized the PSFs with a power-two super-Gaussian with a FWHM of 33 $\lambda/D$, placed at the location of the interferogram.
This location is precomputed according to the plate scale of the data. 
The pipeline builds a fringe library for every interferogram at its own location, such that a direct fit can be made without shifting any interferograms. 
The fit is a least-squares optimization with a model matrix that contains the flattened versions of all fringes in the fringe library.
  
The Fourier method uses the same super-Gaussian masks for both components.
Masking causes information from neighboring $uv$-pixels to blend, such that the central pixel value is representative of the entire splodge. 
Visibilities are then extracted at the central locations of splodges in the $uv$-plane.
For the holographic component, we masked individual interferograms before doing the Fourier transform. 
It is not necessary to center on the interferograms. 
While this does introduce a large phase slope in the $uv$-plane, we know the location of the PSF and can subtract a precomputed phase slope.
Moreover, if we assume that there is no distortion or only symmetric distortion in the image plane, we can average the visibilities of the interferograms with opposite circular polarization.
As they are on exactly opposite sides of the image center, the phase slopes cancel each other out.}

\subsection{Closure phases }
\begin{figure}
    \centering
    \includegraphics[width=0.45\textwidth]{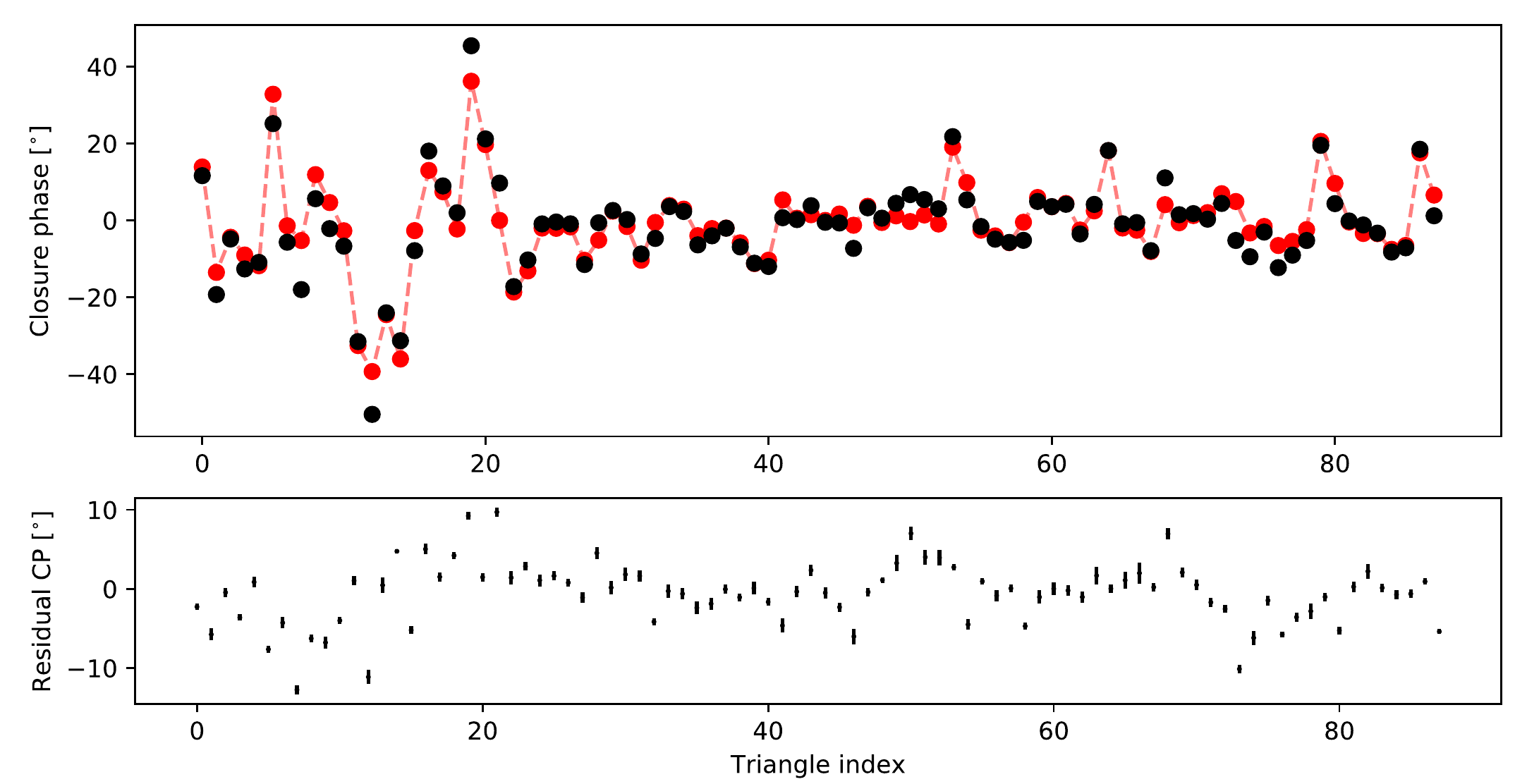}
    \caption{\rch{Comparison of the measured and simulated closure phases of the central component. \textit{Top:} Retrieved closure phase of each closure triangle at 1400 nm (black)}, in addition to closure phases of a forward model (red). The forward model includes a linear polarization fraction of the source,  polarization leakage, and low-order aberrations. \textit{Bottom:} Residuals between the model and the observed closure phases. The error bars indicate the standard deviation of the residual closure phases over 22 hours of measurements at 15 minute intervals. } 
    \label{fig:HAM_CC_CP}
\end{figure}

We extracted closure phases of both the central component and the holographic component of the full HAM PSF using our pipeline.
In Fig. \ref{fig:HAM_CC_CP} we show the closure phases of the central component for 1400 nm, some with large deviations from zero. 
As explained in Sect. \ref{sec:leakage_influence} and in Appendix \ref{app:leakage}, these closure phases are nonzero due to the sensitivity to wavefront aberrations and the polarization state of the incoming light. 
Therefore, we fit a simple model that includes some low-order aberrations, a linear polarization fraction, and polarization leakage to the closure phases.
The low-order aberrations are represented with ten Zernike modes starting with defocus.
A monochromatic PSF is calculated using \texttt{HCIPy} with matching plate scale to the laboratory PSF, and our pipeline extracts the closure phases from the PSF. 
The best fit is shown in red in Fig. \ref{fig:HAM_CC_CP} and is capable of explaining almost all the features of the data. 
The model has a linear polarization fraction of 7\%, a polarization leakage of 3\%, and a 47 nm root-mean-square (RMS) wavefront error, consisting mostly of defocus (38 nm RMS), astigmatism (16 nm RMS), and coma (20 nm RMS). 
Significant residuals remain, which could be explained by effects not present in our simple model, such as higher-order aberrations, inaccurate sampling of the model PSF, or fast-axis deviations from $\pi/2$ in the phase-shifted subapertures. 

The nonzero closure phases again showcase the sensitivity of HAM to aberrations and the polarization state of the incoming light.
When the polarization state and wavefront aberrations are stable in time, the closure phases should stay constant. 
We address the stability in the laboratory by imaging the PSF 90 times over 22 hours, roughly 15 minutes apart.
The camera will not return to exactly the same location on the 5x5 grid for every mosaic. 
Shifts in these positions bias the stability measurement significantly, inflating the error bars.
To be less dependent on the registration of translation between images in the mosaic, we first registered the translation between all images of a single camera position.
We minimized the difference between these 90 images by aligning them all to the first frame in time.
After all images of a single camera position are aligned with respect to one another, we repeated the process for all other camera positions.
Then, we calculated the translation of the 25 camera positions with the first frames of the aligned images.
Because all images are aligned, this offset is the same for all 90 mosaics.
\begin{figure}
    \centering
    \includegraphics[width=0.45\textwidth]{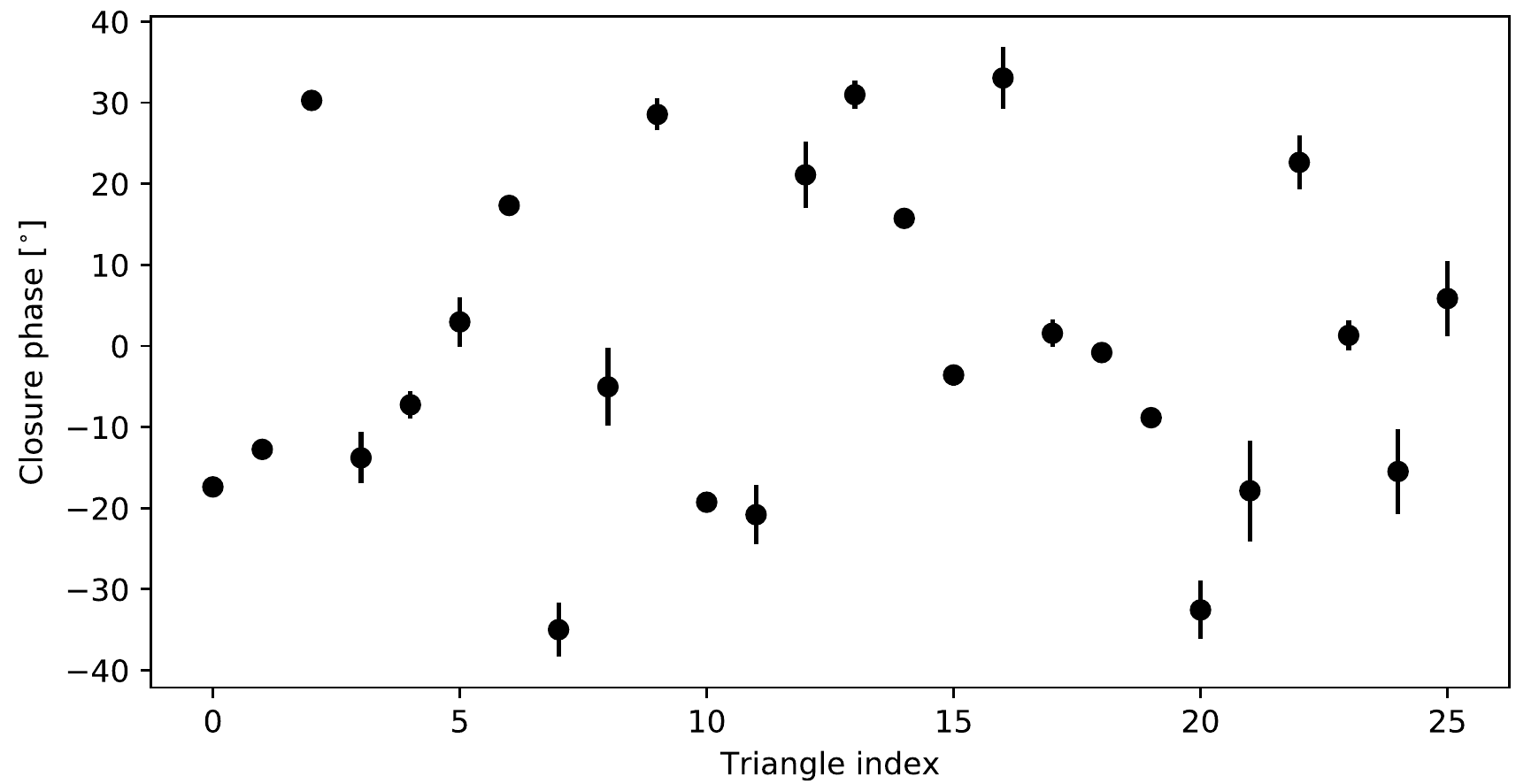}
    \caption{Measured average closure phases of the holographic component over 22 hours of measurements with 15 minute intervals.}
    \label{fig:HAM_HC_CP}
\end{figure}
We generated the 90 mosaics by translating the aligned images with this general offset and stitched them together like before. 

From the PSF in each mosaic, we extracted the closure phases of the central component, and the standard deviation of each closure phase is indicated by the error bars in the bottom plot of Fig.~\ref{fig:HAM_CC_CP}. 
Almost all closure phases have a standard deviation of less than one degree, indicating that they are stable under laboratory conditions. 
This is also true for the underlying cause of the offsets, that is, the polarization fraction and the wavefront aberrations. 
Additionally, we extracted the closure phases of the holographic component and calculated the time variability. 
The results are presented  in Fig. \ref{fig:HAM_HC_CP}, which shows two interesting features.
First, there is a deviation from zero for these closure phases as well, and it is independent of the number of interferograms per closure phase. 
These closure phases are not affected by polarization leakage as the holographic component is imaged far away from the central component.
Second, there is a difference in time variability, for example, the largest variation is 6 degrees compared to less than 1 degree for the first three closure triangles. 
The most stable closure triangles are the ones imaged in a single interferogram (i.e., closure triangles 0, 1, 2, 10, 14, and 18 in Fig. \ref{fig:HAM_HC_CP}). 
This suggests that the time variability does depend on the number of interferograms per closure phase. 
Moreover, it suggests that the cause of the closure phase offset is different than the cause of the variability differences. 
Most likely, the difference in variability is not caused by effects of the optical setup as this would have impacted the closure phases of the central component as well.
A possible explanation for the variability difference could be that there are still remaining translation registration errors for different interferograms in time. 

\rch{On the other hand, the offsets from zero for the closure phases of the holographic component could be generated by differential aberrations in the optical system. 
Light from each subaperture has a different optical path between the HAM optic and the focal plane.
In the laboratory setup, the diffraction angles were large and the imaging doublet could introduce aberrations that change with imaging location on the detector.}
Therefore, the fringe shift can differ for each baseline, such that combining the phases into a closure phase does not add up to zero.
Yet, as the optical setup itself is static, they are stable in time.
%

\section{On-sky verification with HAM v1}
\label{sec:on-sky}
We observed the binary HD 90823 (also known as HDS 1507 or WDS 10294+1211) and an unresolved reference star in two different filters. 
The primary goal was to verify whether the correct system parameters of HD 90823 (i.e., contrast ratio and separation) can be inferred from the data. 
Moreover, we aimed to assess the broadband performance of HAM by extracting wavelength-dependent closure phases from the holographic spots in the focal plane.
In the simple case of a binary, the apparent angular separation between the two stars should not change as a function of wavelength.   
\subsection{Binary HD 90823}
HD 90823 is an ideal verification target because (i) the binary has a relatively low contrast, $\Delta m \approx 1.2$, in the $V$ and $I$ bands, making the companion easy to detect and (ii) two sets of orbital elements have been published in the literature (\citealt{Cvetkovic2016}, \citealt{Tokovinin2017}), allowing us to compute the predicted on-sky separation vector at any point in time. 
In 2016, \citeauthor{Cvetkovic2016} published a result that was based on four measurements spread over more than a decade. 
\rch{The authors found an orbital period of 23 years, yet they warned that their result is \rch{"highly tentative."}
Based on three new data points, \citeauthor{Tokovinin2017} revised the orbital elements in 2017 and lowered the binary's period to just over 15 years. }
Table \ref{tab:binay_observables_literature} provides a further overview of the relevant HD 90823 parameters that are reported in both papers.

\begin{table}[h!]
        \centering 
        \captionsetup{justification=centering}
        \caption{Overview of different HD 90823 parameters as reported in the literature. }
        \vspace{-5pt}
        \begin{threeparttable}
                \begin{tabular}{l c c}
                \hline \hline
                        & \citet{Cvetkovic2016} \ & \citet{Tokovinin2017} \\ 
                        \hline
                        Period (years) & 23.361 & 15.59 $\pm$ 0.11 \\
                        M$_{A}$ ($M_\odot$) & 1.66 & 1.53 \\
                        M$_{B}$ ($M_\odot$) & 1.30 & 1.18 \\
                        Contrast ($\Delta m$) & 1.19 $\pm$ 0.17 ($V$) & 1.21 $\pm$ 0.15 ($I$) \\
                        Spectral type & F0 (A) + F7 (B) & F2 \\
                        \hline
                \end{tabular}
          \begin{tablenotes}
      \small
      \item A and B refer to the binary components. For further details, please refer to the tables in the cited papers.
    \end{tablenotes}
        \end{threeparttable}
        
        \label{tab:binay_observables_literature}
\end{table}

\subsection{Observations} 
\label{sec:observations}

The on-sky verification of the prototype HAM v1.0 test took place on the evening of April 16, 2019 (Hawaiian time). 
Observations were carried out in the $H$ band, and time was divided equally among the binary HD 90823 and the calibrator source HD 90700.
According to the SIMBAD database, the difference between the apparent magnitudes of both objects in the $H$ band is small: $m_H = 6.2$ for HD 90823 versus $m_H = 5.6$ for HD 90700. 
Both objects were observed in a narrowband filter (henceforth Hn5) and a broadband filter with a 20\% bandwidth (henceforth Hbb. 
An observing log and the details of the filters are provided in Tables \ref{tab:observing_log} and \ref{tab:filter_details}, respectively.

The OSIRIS Imager comprises a Teledyne Hawaii-2RG HgCdTe detector with a size of 2048$\times$2048 pixels and a minimum integration time of 1.476 seconds (see \citealt{Arriaga2018} for further specifications). 
Because this number exceeds the typical time associated with atmospheric seeing and the subapertures are larger than $r_{0}$ at the filter bandwidths, aperture masks in OSIRIS can only operate in conjunction with AO.\begin{figure}
    \centering
    \includegraphics[width=0.45\textwidth]{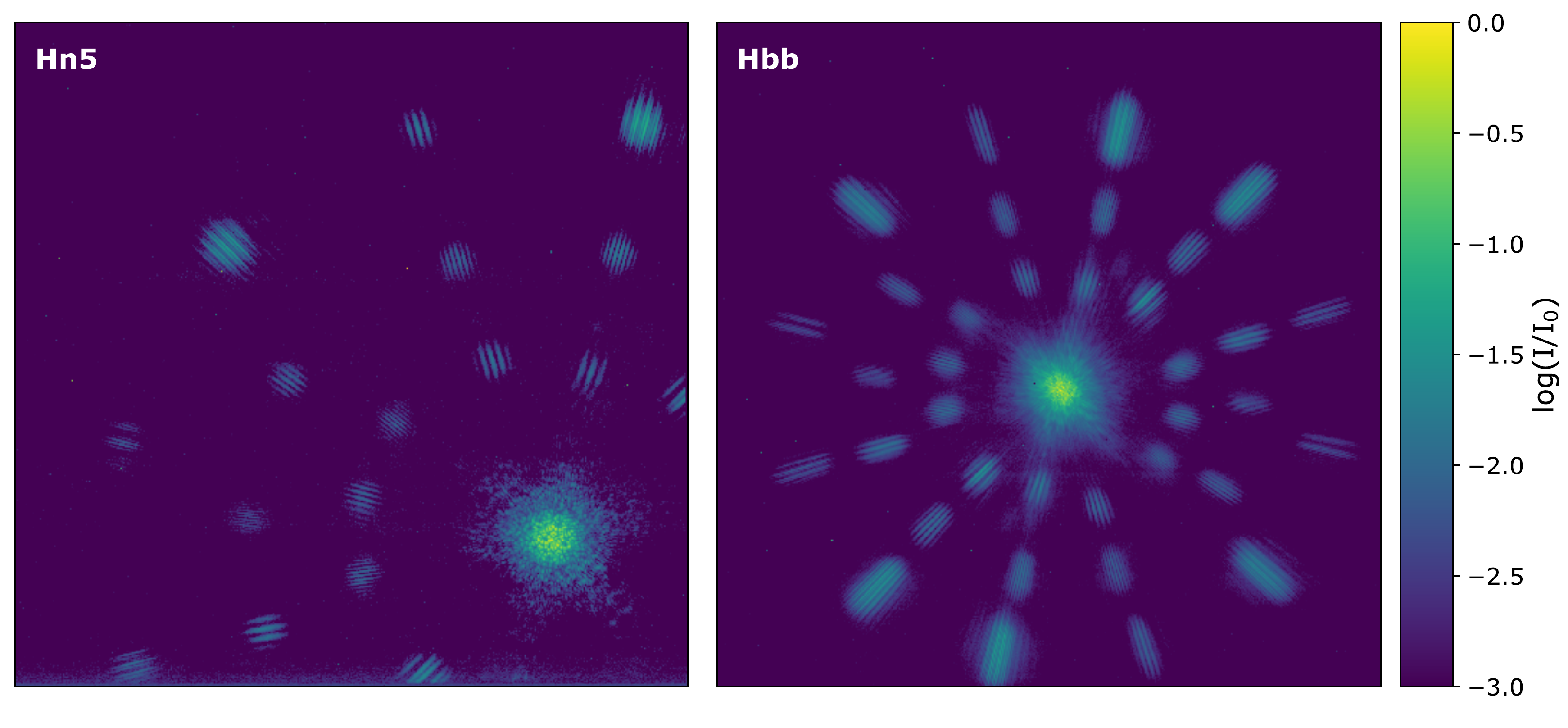}
    \caption{Star HD 90700 imaged with HAM in the Hn5 filter (left panel) and the Hbb filter (right panel). These figures only display a sub-window of the full sensor area. Data recovered for the Hn5 PSF occupied the bottom-right corner of the detector, so not all holographic spots are visible. \rch{The Hn5 image is cropped for clarity.}}
    \label{fig:HAM_PSF_onsky}
\end{figure}
\begin{table}[h!]
        \centering 
        \captionsetup{justification=centering}
        \caption{Summary of the observations taken with HAM at Keck on the evening of April 16, 2019 (Hawaiian time).}
        \begin{threeparttable}
                \begin{tabular}{c c c c c c} 
                \hline 
                \hline
                        Start (UTC) \ & End (UTC) \ & Target & Filter & N$_{f}$ & t$_{exp}$ (s) \\ 
                        \hline
                        06:28:52 & 06:36:30 & 90700 & Hn5 & 50 & 1.476 \\
                        06:39:05 & 06:46:42 & 90823 & Hn5 & 49\tnote{$a$}  & 1.476 \\
                        06:49:48 & 06:57:26 & 90823 & Hbb & 48\tnote{$b$}  & 1.476 \\
                        06:58:56 & 07:06:47 & 90700 & Hbb & 50 & 1.476 \\
        \hline
                \end{tabular}
                \begin{tablenotes}
                        \item[$a$] {\small \sffamily One corrupted frame was discarded.}
                        \item[$b$] {\small \sffamily Two frames with a much lower signal-to-noise were discarded.}
                \end{tablenotes}
        \end{threeparttable}
        \label{tab:observing_log}
\end{table} 
\begin{table}[h!]
        \centering 
        \captionsetup{justification=centering}
        \caption[]{Properties of the two filters that were used during the observations.}
        \vspace{-5pt}
        \begin{threeparttable}
                \begin{tabular}{@{}l c@{} c@{} c@{} c@{} c@{}}
                \hline \hline
                        \rch{Filter} & PSF position & $\lambda_0$ (nm) \ & $\Delta \lambda/\lambda_0$ (\%) \ \\
                        \hline
                        \rch{Hn5} & Bottom-right \ & 1765 & 4.9\% \\
                        \rch{Hbb} & Center  & 1638 & 20.1\% \\
                        \hline
                \end{tabular}
        \begin{tablenotes}
      \small
      \item Values were copied from the OSIRIS filter table on the Keck website\footnotemark.
    \end{tablenotes}
                 
        \end{threeparttable}
        \label{tab:filter_details}
\end{table}
\footnotetext{\url{https://www2.keck.hawaii.edu/inst/osiris/scale_filter.html}}
Tests with an internal source showed that the imaging quality changed considerably as a function of the input source location in the field of view (FOV). 
The separation between the HAM phase and amplitude mask of HAM v1.0 was considered as the source of this changing imaging quality.
Light that passes through the holes of the amplitude mask at an angle to the optic axis then intercepts the phase mask off-axis, possibly illuminating the phase mask beyond the edges of the phase pattern of individual holes.
If present, such a leakage term would affect the central component PSF.
However, images of some sections of the HAM mask pupil did not show partial illumination of holes. 
As the whole optic could not be imaged in pupil viewing mode, we cannot be certain that all holes were fully illuminated.
Another explanation could be a differential focus between the OSIRIS Imager and the OSIRIS spectrograph, which was present at the time of observation. 
Reconstruction of the visibility amplitudes also showed a gradient in the pupil illumination.
The OSIRIS Imager was realigned, and HAM v1.5 was installed after the presented observations. 
New internal source measurements with the HAM v1.5 mask show little variation of the PSF quality as a function of the position in the FOV. 

Because the image quality of the central component looked highest in the bottom-right corner of the detector, we decided to locate it there during the Hn5 observations (sacrificing the majority of holographic spots; see the left panel of Fig. \ref{fig:HAM_PSF_onsky}).
However, with the holographic spots being of primary interest in the broadband, the PSF was shifted to the middle of the detector for the Hbb observations (right panel of Fig. \ref{fig:HAM_PSF_onsky}). 
This allowed us to compute HAM's full set of closure phases, albeit with lower image quality.    
\subsection{Data reduction} \label{sec:data_reduction_ch4}
We applied a dark correction using 100 dark frames and calculated the power spectrum of the holographic component and the central component separately on a $uv$-grid of 3000$\times$3000 pixels. 
For the Hn5 filter we masked everything but the central component, while for the Hbb filter we applied the same data reduction as mentioned in Sect. \ref{sec:pipeline}.
Before the fringe phases could be extracted, we needed to determine exactly where in the $uv$-plane the visibility had to be sampled. 
This required two pieces of information: (i) the rotation angle of the PSF, which is set by the mask's orientation with respect to the detector, and (ii) the radial scaling of the PSF, which depends on the wavelength and the magnification of the instrument. 
To find the optimal parameter values that describe the scaling and orientation of the power spectrum (and thus of the PSF), we used a model of the $uv$-plane that is cross-correlated with the observed power spectra as a function of radial scaling and rotation angle. 
We found that the angular orientations of the power spectra of the holographic and central components differ by $\sim 2.1^\circ$. 
This suggests that there is an angular offset between the amplitude mask and the phase mask in both filter wheels. 
The orientation of the central component's power spectrum is determined by the holes in the amplitude mask, while the orientation of the holographic component's power spectrum is determined by the slopes on the phase mask, which has to be taken into account when fitting a model to the observed closure phases.
We extracted closure phases from the observed PSFs using the Fourier methods for both monochromatic and broadband observations.
\subsubsection{Closure phases} 
Once all visibility phases were sampled in the $uv$-plane, we computed the closure phases associated with each triangle on the mask. 
This resulted in 197 sets of closure phases, each of which corresponds to one of the science frames listed in Table \ref{tab:observing_log}.
In addition, we averaged the closure phases of the calibrator HD 90700 per filter, and we considered the standard deviation as the measure of the stability of a closure triangle over time.
We then calibrated the binary closure phases of each frame by subtracting the average closure phases of the calibrator. 

\begin{figure}
    \centering
    \includegraphics[width=0.5\textwidth]{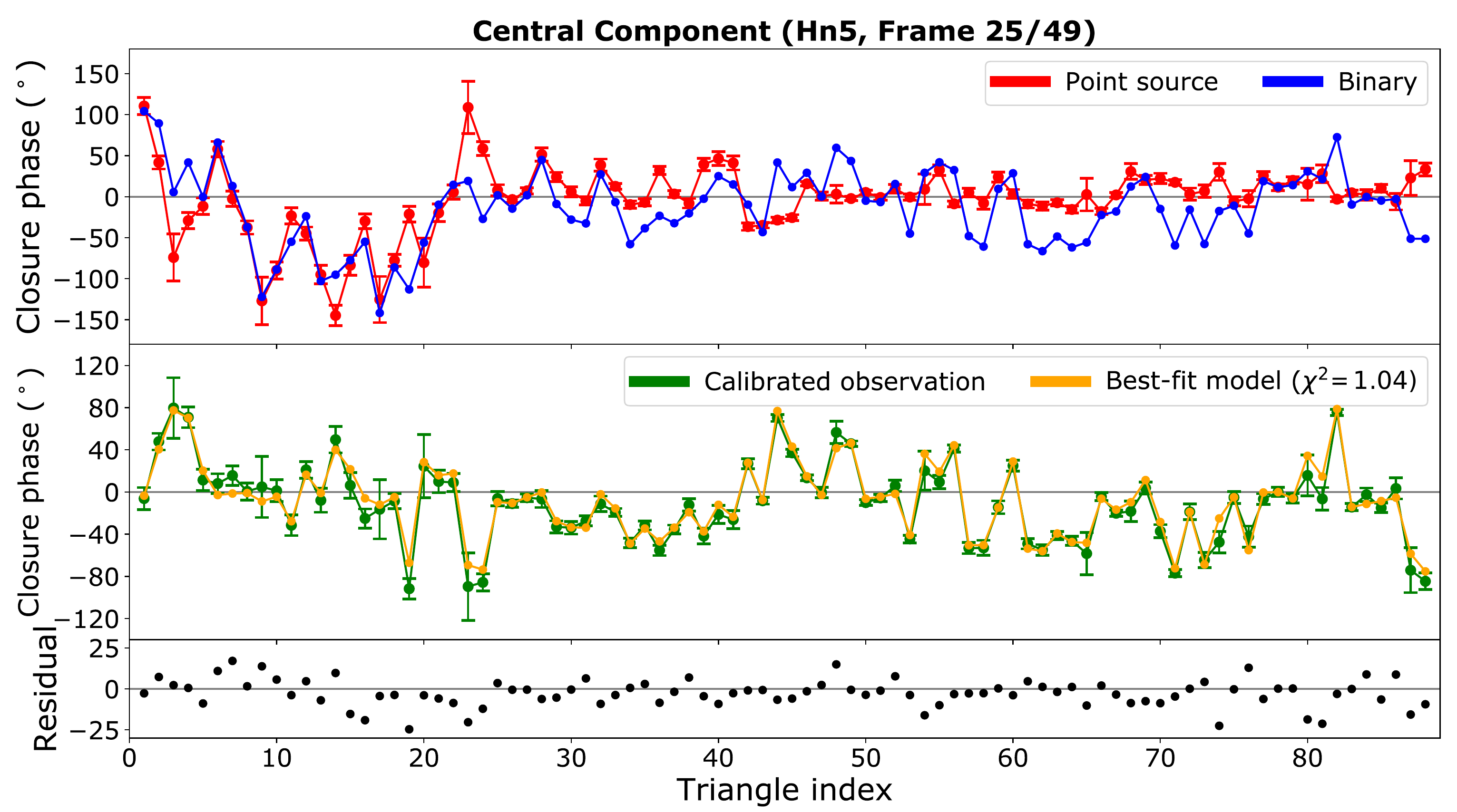}
        \caption{\rch{Model fit to the calibrated closure phases of the HD 90823 (binary) system.} \textit{Top:} Closure phases of HD 90823 inferred from the central component in \rch{frame 25 (out of 49 frames)} in the Hn5 filter. The closure phases of HD 90700 (point source) are averaged over all frames. \textit{Middle:} Calibrated observation, obtained by subtracting the closure phases of the point source from the closure phases of the binary. The best-fit model that was found is overplotted, with the corresponding $\chi^2$ value reported in the legend. \textit{Bottom:} Residuals left after subtracting the model from the observation.} 
    \label{fig:HAM_CP_Hn5}
\end{figure}

\begin{figure}
    \centering
    \includegraphics[width=0.5\textwidth]{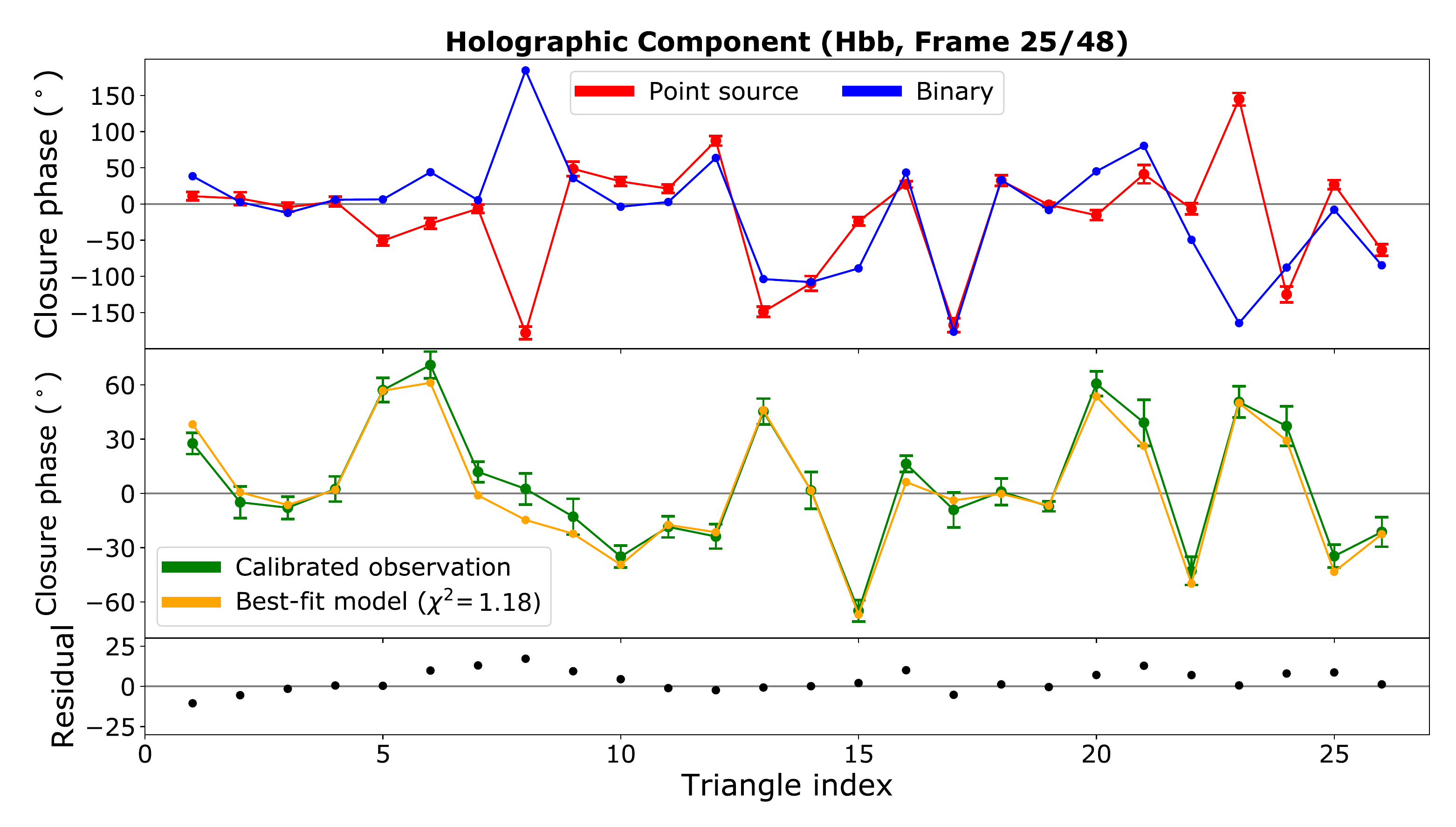}
        \caption{Same as Fig. \ref{fig:HAM_CP_Hn5}, but now for the \rch{holographic spots} in \rch{frame 25} in the Hbb filter. The $uv$-plane was sampled at 1638 nm.}
    \label{fig:HAM_CP_Hbb}
\end{figure}

Some representative results are displayed in Figs. \ref{fig:HAM_CP_Hn5} and \ref{fig:HAM_CP_Hbb}. 
Figure \ref{fig:HAM_CP_Hn5} shows both the non-calibrated and calibrated closure phases of HD 90823 obtained from the PSF core \rch{in the 25th of 49 frames} in the Hn5 filter, as well as the best-fit model that was found. 
The average closure phases of the point source are plotted in the background.
From Fig. \ref{fig:HAM_CP_Hn5} it is clear that the closure phases of HD 90700 (point source) exhibit a remarkably strong deviation from zero.
The deviations are much stronger for HAM v1.0 on-sky than the offsets measured in the laboratory for HAM v1.5, as presented in the previous section.
It is unclear if this is purely due to the differences between HAM v1.0 and HAM v1.5.
However, the overall structure of the closure phases is in agreement with the laboratory measurements, with the first 24 closure phases (with triangle indices $\leq$ 24) of the central mask component having the largest deviations.
Moreover, Fig.~\ref{fig:HAM_CP_Hn5} suggests that the errors on these 24 triangles are also greater than the other errors as they are more sensitive to changes in defocus. 
Even the PSF at the corner of the detector in the Hn5 filter has a poor quality, indicating that the PSF is strongly aberrated.
Even with small fractions of linearly polarized light present, it is expected that the closure phase offsets are much larger under these conditions. 
This implies that calibration is essential for extracting physical information from the data.

Figure \ref{fig:HAM_CP_Hbb} shows the measured closure phases \rch{for the 25th of 48 frames} in the Hbb filter for the holographic spots at the central wavelength, $\lambda_0$. 
Again, the closure phases for the single star deviate strongly from zero and, again, much more than the laboratory measurements. 
We did not find a good explanation for this offset, but we can calibrate it for the binary system using the calibrator.

\subsubsection{Parameter estimation}
Given the measured closure phases and calibration of the zero points, the separation ($\boldsymbol{\rho} = (\rho_x, \rho_y)$) and the contrast ratio of HD 90823 can be estimated. 
We did this by finding the parameter combination $(r, \boldsymbol{\rho})$ that minimizes the chi-squared difference, $\chi^2$, between the observation and an analytical model of a binary system:

\begin{equation} \label{eq:chi_squared_observations}
\chi^2(r, \boldsymbol{\rho}) = \frac{1}{N-m} \sum_{i=1}^{N} \Bigg( \frac{[ \phi_{i, \rm obs} - \phi_{i, \rm mod}(r, \boldsymbol{\rho}) + \pi ] \% (2\pi) - \pi}{\sigma_i} \Bigg)^2.
\end{equation}

Here, $N$ is the number of closure phases, m is the number of free parameters, $\phi_{i, \rm obs}$ is the $i$-th observed closure phase (in radians, with error $\sigma_i$), and $\phi_{i, \rm mod}$ is the $i$-th modeled closure phase (in radians).
Adding $\pi$, applying the modulo operator percentage, and subtracting $\pi$ ensures that all differences are mapped onto the interval [$-\pi$, $+\pi$]. 
We performed a total of 145 fits: 49 in the Hn5 filter and 96 (2$\times$48) in the Hbb filter, whereby the closure phases obtained from the PSF core and the holographic spots were treated separately. 
\begin{table}[h]
        \centering 
        \captionsetup{justification=centering}
        \caption{Estimated separation and contrast ratio for HD 90823.}
        \vspace{-5pt}
        \begin{threeparttable}
                \begin{tabular}{@{}l c@{} c@{} c@{}}
                \hline \hline
                        Subset & N$_f$ \ & Separation $\rho$ (mas) \ & Contrast ratio $r$ \ \\ \hline
                        Central (Hn5) & 38 & 121.9      $\pm$ 0.5 & 0.45 $\pm$ 0.01\\
                        Central (Hbb) & 47 & 121.1      $\pm$ 0.8 & 0.45 $\pm$ 0.02\\
                        Holographic(Hbb) & 46 & 120.9  $\pm$ 0.5 & 0.44 $\pm$ 0.02 \\
                        Full HAM (Hbb) & 93 & 121.0 $\pm$ 0.7 & 0.44 $\pm$ 0.02\\ \hline
                \end{tabular}
    \begin{tablenotes}
      \small
      \item The closure phases in the broadband Hbb filter were sampled at the central wavelength, $\lambda_0$.
    \end{tablenotes}
                
        \end{threeparttable}
        \label{tab:results_binary_parameters}
\end{table}
Table \ref{tab:results_binary_parameters} provides an overview of the HD 90823 parameters that were found after evaluating Eq. \ref{eq:chi_squared_observations} on a high-resolution grid. 
The reported values for $\rho = \sqrt{\rho_x^2 + \rho_y^2}$ and $r$ result from averaging the best-fit parameters over different subsets of frames. 
Only fits for which $\chi^2 < 2N_{CP}$ (where $N_{CP}$ is the number of closure phases) are included to reduce the effect of outliers. 
We find the same contrast ratio, $r = 0.45 \pm 0.02$, for each of the different subsets. 
The value inferred for the separation, $\rho$, is \textasciitilde1 mas larger in the Hn5 filter as compared to the Hbb filter, which can only be explained by random errors. 

As far as the performance of the mask's holographic component is concerned, it is reassuring that the retrieved parameter values in the Hbb filter are consistent with each other.
The observables obtained from the central and holographic components of the mask are within $1\sigma$. 
\subsection{Spectroscopic parameter retrieval}
\begin{figure}
    \centering
    \includegraphics[width=0.5\textwidth]{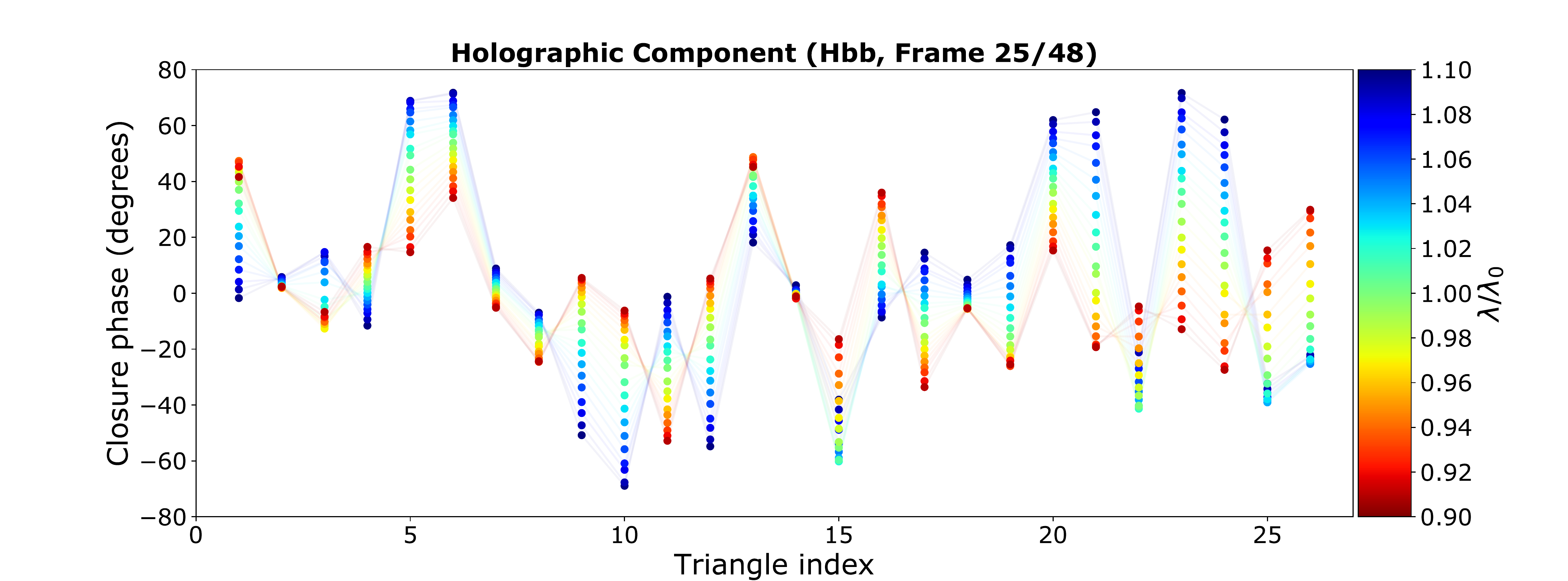}
        \caption{Calibrated closure phases obtained from the \rch{holographic spots} in \rch{frame 25} in the Hbb filter, plotted for different wavelengths within the 20\% bandwidth (see color bar).}
    \label{fig:HAM_CP_broadband}
\end{figure}

Holographic aperture masking has the unique capability to extract low-resolution spectroscopic closure phases using the holographic component.
Here, we look at the multiwavelength extraction of the closure phases, which allows us to extract the separation and magnitude from multiple wavelength channels in the Hbb band, spanning from 1473 nm to 1803 nm.
We illustrate the wavelength dependence of the measured closure phases in the Hbb filter in  Fig. \ref{fig:HAM_CP_broadband}. 
The closure phases from the holographic spots are sampled at different coordinates in the $uv$-plane (corresponding to different wavelengths).
The number of sampling points far exceeds the number of independent measurements.
\rch{The spectral resolution will be two or three}, depending on the triangle index, according to Eqs. \ref{eq:spec_res1} and \ref{eq:spec_res2}.
The closure phases behave well, and there seems to be a smooth transition from one wavelength to another. 

\begin{figure}
    \centering
    \includegraphics[width=0.5\textwidth]{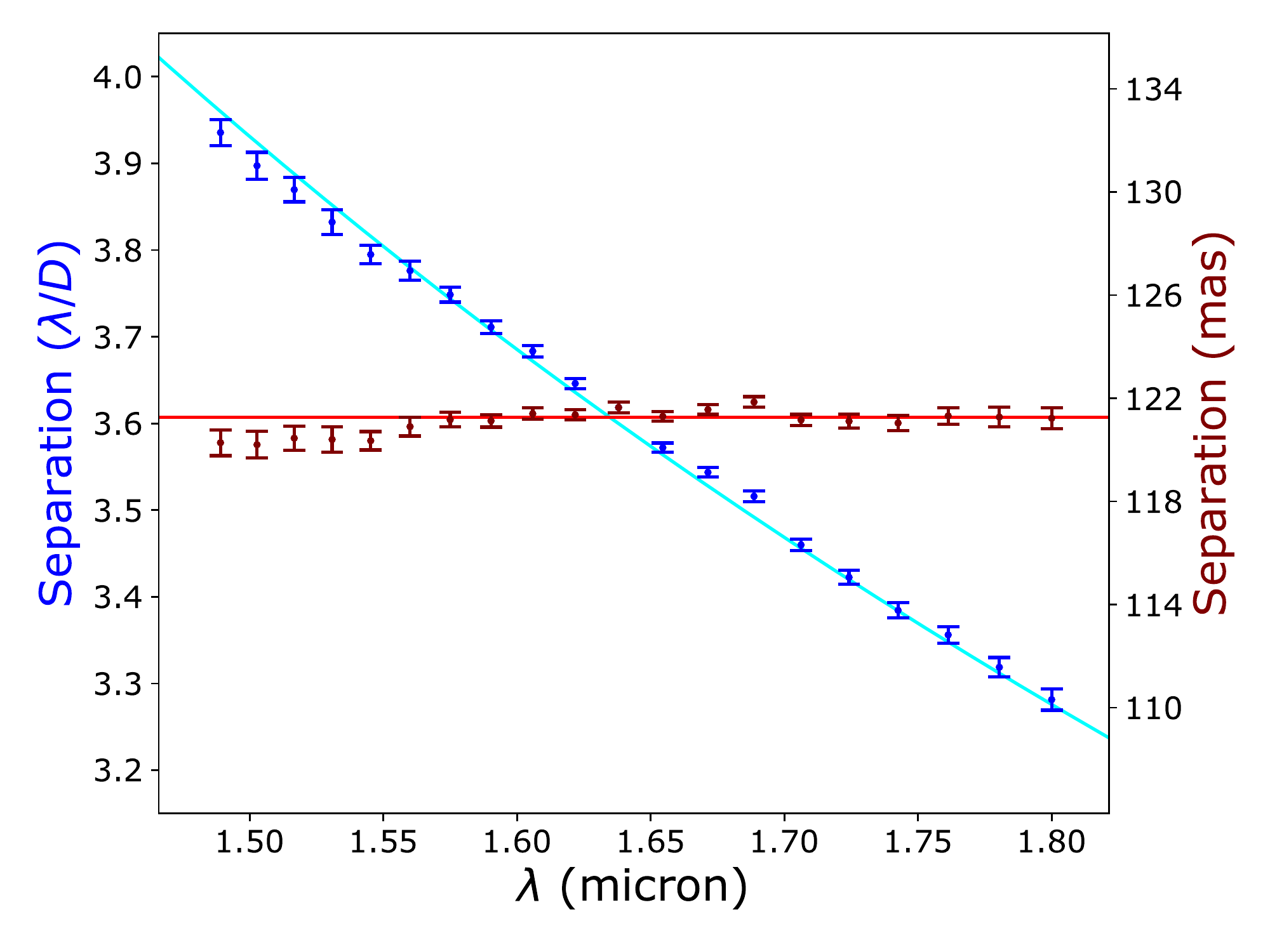}
        \caption{Separation between the components of HD 90823 inferred with HAM as a function of wavelength, expressed in units of $\lambda/D$ (blue points) and expressed in milliarcseconds (red points). The blue line is proportional to $1/\lambda$. The red line is a weighted average of the measured separations in milliarcseconds. \rch{The number of data points is larger than the number of independent measurements ($\sim 4$).}}
    \label{fig:binary_sep_wl}
\end{figure}

\begin{figure}
    \centering
    \includegraphics[width=0.5\textwidth]{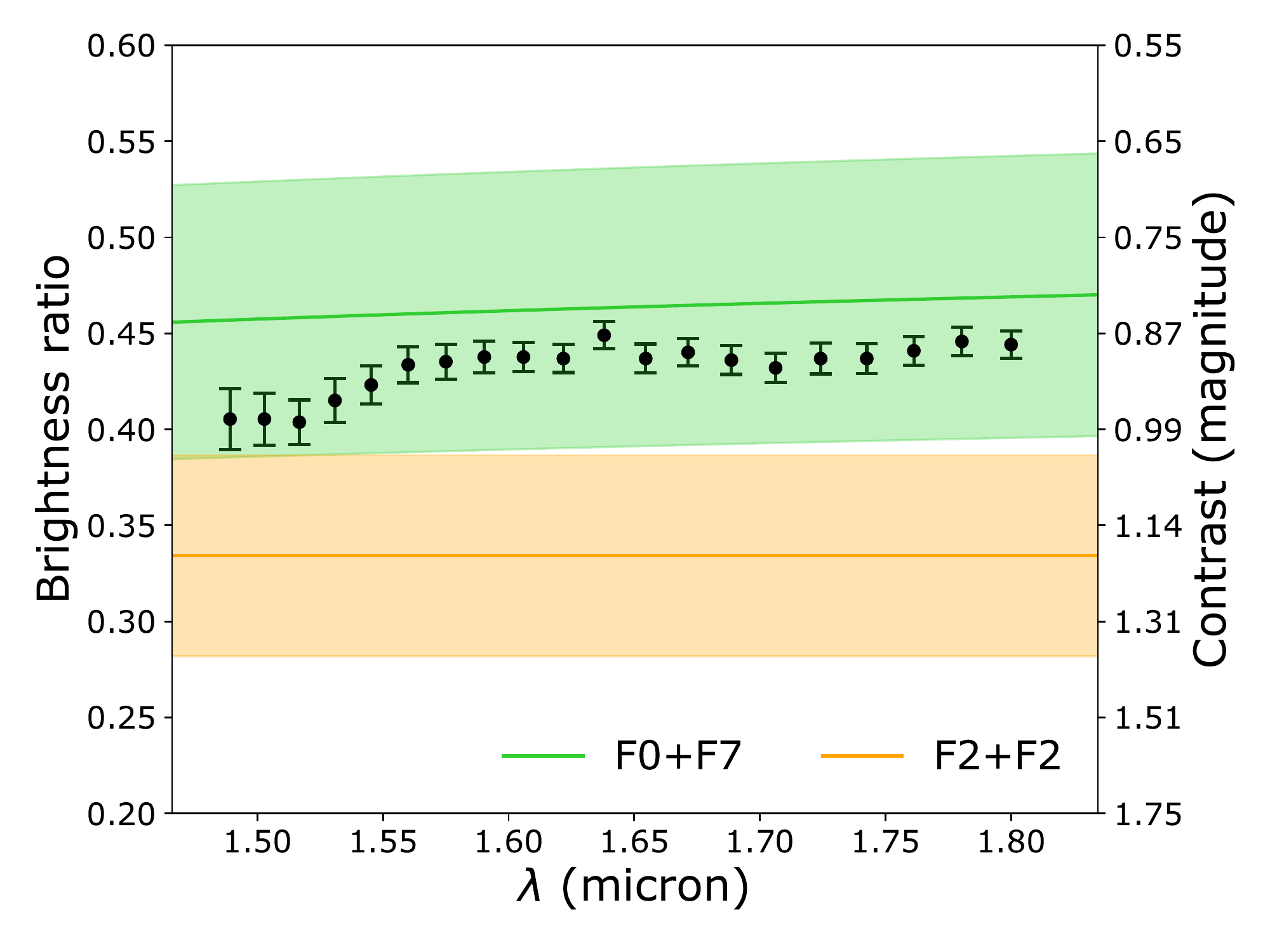}
        \caption{Recovered brightness ratio of the binary as a function of wavelength. The green line shows the expected brightness ratio for the spectral types (F0 and F7) reported by \citet{Cvetkovic2016}. The orange line is the expected brightness ratio when both components are of type F2. The envelopes represent the 1$\sigma$ error from the $V$-band and $I$-band measurements presented in Table \ref{tab:binay_observables_literature}. \rch{The number of data points is larger than the number of independent measurements ($\sim 4$).}}
    \label{fig:binary_con_wl}
\end{figure}

We extracted the binary parameter values by minimizing the $\chi^2$ (Eq. \ref{eq:chi_squared_observations}) for closure phases sampled at different points in the $uv$-plane.
Figures \ref{fig:binary_sep_wl} and \ref{fig:binary_con_wl} display the parameters of HD 90823 that were inferred from the holographic spots in the Hbb filter as a function of wavelength. 
Figure \ref{fig:binary_sep_wl} shows the separation, $\rho$, between the binary components. 
Expressed in units of $\lambda/D$, the separation exhibits a $1/\lambda$ drop-off, which implies that $\rho$ must be constant throughout the bandwidth. 
This is the expected result for a binary system, and it is a powerful method for distinguishing astronomical observables from instrument artifacts. 
As mentioned in Table \ref{tab:results_binary_parameters}, we find a separation of 121$-$122 mas based on the closure phases at $\lambda_0$. 
This is also the value that follows from averaging over all wavelengths in the bandwidth, as shown by the horizontal line. 
Some points deviate significantly from the average.
The error bars are determined using a jackknife method \citep{roff1994}, which does not take systematic errors into account (e.g., wavelength-dependent errors in closure phase retrieval).
\noindent The measured brightness ratio, $r$, as a function of wavelength is plotted in the right panel of Fig.  \ref{fig:binary_con_wl}. 

\begin{figure*}
    \centering
    \includegraphics[width=0.85\textwidth]{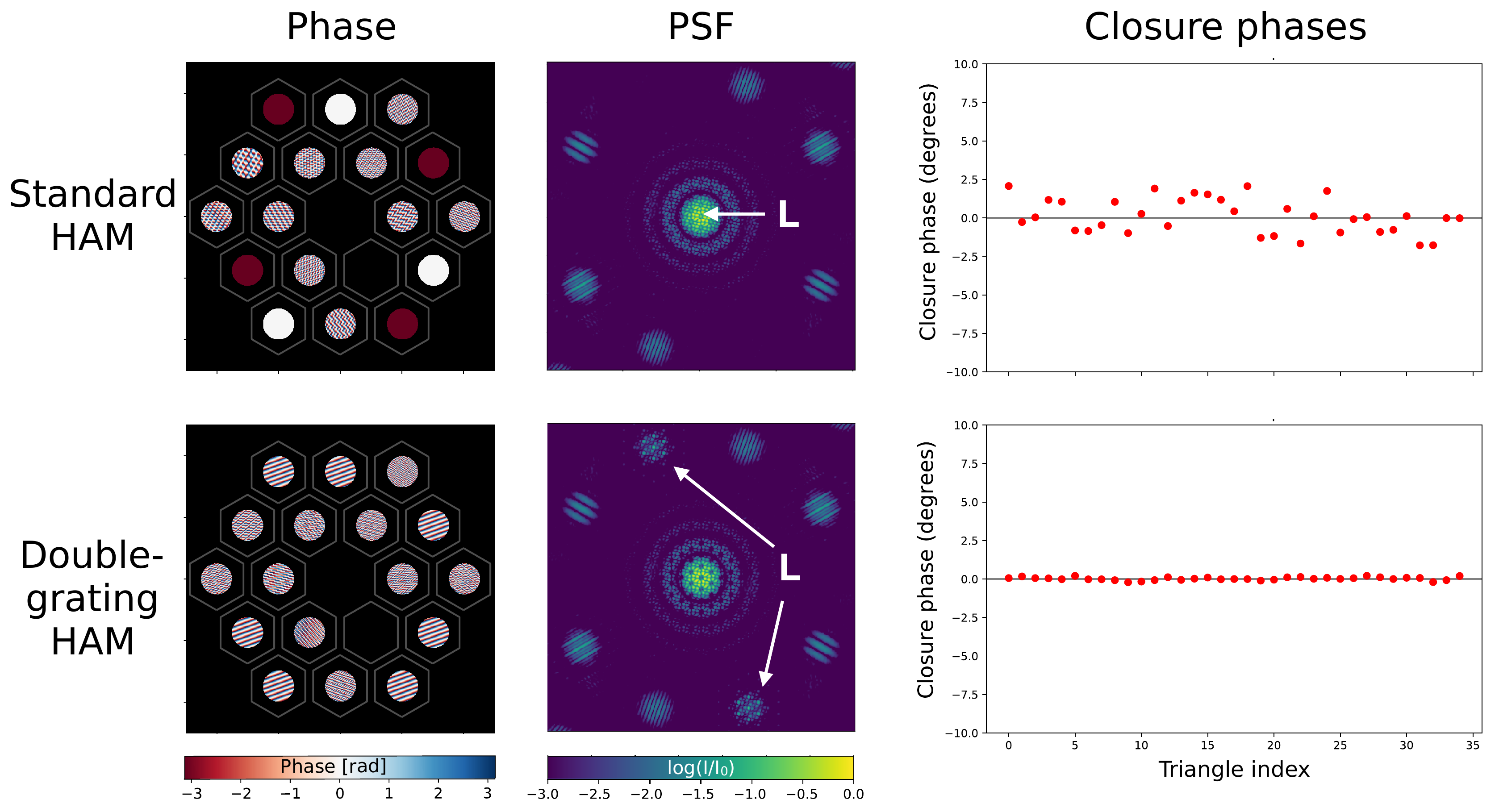}
        \caption{Difference between a standard HAM and a double-grating HAM. The first element of the double grating is the standard HAM pattern with an added polarization grating pattern \rch{(i.e., the phase ramp)} with 70 periods across the full pupil. The second grating is not shown. The resulting PSFs show that the polarization leakage is directed from the center into two off-axis PSFs. Both elements have 2\% leakage, and we assume a linear polarization fraction of 10\%. Closure phase offsets due to polarization leakage are greatly reduced with the double-grating method.}
    \label{fig:double-grating-HAM}
\end{figure*}
The spectral types of the HD 90823 components are uncertain.
According to the SIMBAD database, both stars are of type F2, but the corresponding quality labels suggest little reliability. 
On the other hand, \citet{Cvetkovic2016} state that the bright component is hotter (type F0) than the faint one (type F7), with the difference in surface temperatures being roughly 1000 K. 
In order to find out which claim is most likely based on the HAM data, we computed the Planck spectra of the components in both scenarios, scaled them according to the reported contrast in the$V$ band (see Table \ref{tab:binay_observables_literature}), and divided them in the Hbb filter. 
Figure \ref{fig:binary_con_wl} illustrates that the actual measurements lie in between the theoretical curves but fully within the 1$\sigma$ envelope \rch{of the combined F0 and F7 scenario.}
Moreover, the gradient as a function of wavelength is consistent with the combined F0 and F7 scenario.
This means that our estimate of the brightness ratio also suggests a difference between the \rch{components'} temperatures and spectral types, in line with \citet{Cvetkovic2016}.   

\section{Conclusions and outlook}
\label{sec:conclusions}
In this paper we demonstrate that HAM shows promise to empower a new generation of SAM experiments: retaining the instrumental simplicity while adding spectroscopic capabilities, higher throughput, and added Fourier coverage. We discuss the trade-offs of a prototype HAM mask in OSIRIS. A first version of this mask was installed at Keck OSIRIS in 2018. By observing the binary HD 90823 with this mask, we obtained low-resolution spectra of closure phases, confirming the broadband capabilities of the HAM mask. 
We investigated two limitations of this version. The first was the spatial separation between the phase and amplitude mask, resulting in spatially varying PSF quality. This was solved with an upgraded version, HAM v1.5, with a diced version of the same phase mask combined in the same holder as the amplitude mask, which was installed in OSIRIS in early 2020. Secondly, we found nonzero closure phases for a single star. We showed in simulations that polarization leakage can produce these nonzero offsets in closure phases and confirmed this with laboratory demonstrations.
 
A future upgrade of the HAM mask will be able remove the effects of polarization leakage, by several orders of magnitude suppression, of the unwanted light using the double-grating method \citep{doelman2020minimizing}. 
The double-grating method adds a phase ramp to the phase pattern, so that any polarization leakage travels in a different direction than that of the main beams.
A second grating with the same phase ramp \rch{(i.e., polarization grating)} is installed directly after the first phase pattern, folding the two main beams back on axis.
The polarization leakage term of the first grating, on the other hand, is diffracted away by this second grating, reducing the on-axis leakage by at least an order of magnitude.
The location of the polarization leakage can be controlled with the phase ramp slope and direction, similar to the holographic interferograms. 
Placing the polarization leakage on an empty spot on the detector reduces the phase slope.
Adapting the focal plane design of a HAM device to leave room for this leakage term would be beneficial. 
We simulated the performance of a standard HAM device and a double-grating HAM device, assuming 2\% leakage and a linear polarization fraction of 10\%. 
The results are shown in Fig. \ref{fig:double-grating-HAM}.
Using the double-grating method reduces the standard deviation of the closure phases from 1.1 degrees to 0.1 degree. 
The residual closure phase pattern of the double-grating HAM is not correlated with the standard HAM, which suggests that the deviation from zero is caused by a different effect (e.g., inaccuracies in the data reduction).
These simulations prove that a double-grating version of HAM would greatly reduce the impact of polarization leakage on the HAM performance.

The potential of a double-grating HAM is exciting. 
We outline a few scientific prospects that are enabled only by HAM.
Studies of thermal emission from protoplanets in protoplanetary disks is complicated by disk features that can emulate the exoplanet signals (e.g., light scattered by dust) that these systems can display \citep{Kraus2011, Sallum2015, currie2019no}.
However, spectral information can help with discriminating between disk and planet signals.
The simultaneous spectral and spatial measurements of HAM help constrain infrared spectral slopes, distinguishing between scattered light and thermal signals.
Other opportunities are enabled by the higher throughput of a HAM mask compared to a SAM mask. 
Follow-up on the brightest Transiting Exoplanet Survey Satellite (TESS) targets can help with ruling out background or binary contaminators. 
Monitoring brown dwarf binaries with HAM increases the efficiency of determining orbits and dynamical masses, directly testing predictions for lithium burning, the \rch{stellar-substellar} mass boundary, and substellar cooling rates \citep[e.g.,][]{Dupuy2017}.
Suppressing the polarization leakage will be critical for improving closure phase stability, resulting in better contrast. 

\begin{acknowledgements}
The research of David Doelman and Frans Snik leading to these results has received funding from the European Research Council under ERC Starting Grant agreement 678194 (FALCONER). 
Laser cutting of aperture masks and planar optics was carried out with the assistance of the OptoFab node of the Australian National Fabrication Facility, utilizing NCRIS and NSW Gov. funding. 
We specifically thank Benjamin Johnston from OptoFab for the quick turn-around of the many amplitude masks and for laser cutting the HAM v1.5 mask.
Some of the data presented herein were obtained at the W. M. Keck Observatory, which is operated as a scientific partnership among the California Institute of Technology, the University of California and the National Aeronautics and Space Administration. The Observatory was made possible by the generous financial support of the W. M. Keck Foundation.
The authors wish to recognize and acknowledge the very significant cultural role and reverence that the summit of Maunakea has always had within the indigenous Hawaiian community.  We are most fortunate to have the opportunity to conduct observations from this mountain.
We thank John Canfield and Peter Wizinowich for the technical support.
We also thank Michael Liu for the fruitful discussions, which helped improve the results presented in this work.

\end{acknowledgements}

\bibliographystyle{aa}
\bibliography{HAM}{}

\appendix

\section{Spectral resolution of the holographic interferograms}
\label{app:spectral_resolution}
Here we derive the spectral resolution of holographic interferograms.
We start by looking at a single baseline, $\mathbf{b}$, between two subapertures with a phase ramp with a period P and a direction $\mathbf{\hat{a}}$.
Assuming no piston phase offset between the subapertures, the electric field is given by
\begin{equation}
M(\mathbf{r}) = \Pi(\mathbf{r}) \otimes \left[\left(\delta(\mathbf{r} - \mathbf{b}/2) + \delta(\mathbf{r} + \mathbf{b}/2)\right)\text{e}^{ 2 \pi i \mathbf{ar}}\right],
\end{equation}
where $\otimes$ is the convolution operator, $\delta(\mathbf{r})$ is the Dirac delta function, $\mathbf{a} = P/D_{sub}\mathbf{\hat{a}}$, and $\Pi(\mathbf{r})$ defines the subaperture. 
For a circular subaperture with diameter $D_{sub}$, 
\begin{equation}
 \Pi(\mathbf{x}) = 
    \begin{cases}
      1 & \text{if } |\mathbf{x}| \leq \frac{D_{sub}}{2} \\
      0 & \text{otherwise.}\\
    \end{cases}
\end{equation}
The PSF is then described by
\begin{equation}
p(\mathbf{\theta}) = P(\boldsymbol{\theta},\lambda) \left[2+2\cos{\left(\frac{2\pi\mathbf{b}\boldsymbol{\theta}}{\lambda}\right)}\right]\otimes \delta(\boldsymbol{\theta}-\lambda\mathbf{a}).
\end{equation}
Here, $\boldsymbol{\theta}$ and $\frac{\mathbf{r}}{\lambda}$ are the Fourier plane coordinates, and $P(\boldsymbol{\theta},\lambda)$ is the PSF of the aperture function:
\begin{equation}
P(\boldsymbol{\theta},\lambda) = \text{Airy}\left( \frac{\pi D_{sub} |\boldsymbol{\theta}|}{\lambda}\right)
.\end{equation}
As the PSF location is directly proportional to $\lambda\mathbf{a}$ due to the grating, we have an independent measurement of a baseline phase when the shift is $1.22\lambda/D_{sub}$ (i.e., the Raleigh criterion). 
Therefore, we can define the spectral resolution as
\begin{equation}
R_{fp} = \frac{\lambda}{\Delta \lambda} = \frac{D_{sub}}{1.22 P}.
\label{eq:spec_res1_app}
\end{equation}
Increasing the subaperture size and grating frequency yields a higher spectral resolution.
For a subaperture with $D_{sub} = \frac{1}{10}D$ imaged at 100$\lambda/D$, the period is $D_{sub}/10$ and $R \sim 8$.

The second spectral resolution of the holographic component is defined in the $uv$-plane.
We calculate the $uv$-plane distribution, $\hat{V}(\mathbf{f})$, with the Fourier transform of the PSF,
\begin{equation}
\hat{V}(\mathbf{f}) = \left(\Pi(\mathbf{r})     \star \Pi(\mathbf{r})  \right)  ^{\frac{1}{2}} \otimes \left[ \delta(\mathbf{f} - \mathbf{b}/\lambda) + \delta(\mathbf{f} + 2\delta(\mathbf{f}) + \mathbf{b}/\lambda) \right]\text{e}^{ 2 \pi i \mathbf{af}}.
\end{equation}
We define the cross-correlation between the subaperture function, $Pi(\mathbf{r})$, shifted by  $\mathbf{b}/\lambda$ as a splodge.
The location of the splodges changes with wavelength, and the shift depends on the length of the baseline.
The phases of two splodges of different wavelengths for the same baseline can be uniquely extracted when they are separated by $D_{sub}$.
In principle, this is possible for every SAM mask; however, to increase throughput, many of these masks have large subaperture diameters and many baselines, such that even for a small bandwidth the splodges start to overlap. 
In the case of HAM, the holographic interferograms only contain a limited amount of baselines and the effect of overlapping splodges can be reduced by design.
The spectral resolution in the $uv$-plane is given by
\begin{equation}
R_{uv} = \frac{\lambda}{\Delta \lambda} = \frac{|\mathbf{b}|}{D_{sub}}.
\label{eq:spec_res2_app}
\end{equation}

\section{HAM mask design parameters}
\label{app:HAM_OSIRIS_specs}
\begin{figure}
    \centering
    \includegraphics[trim=0 25 0 15,clip,width=0.5\textwidth]{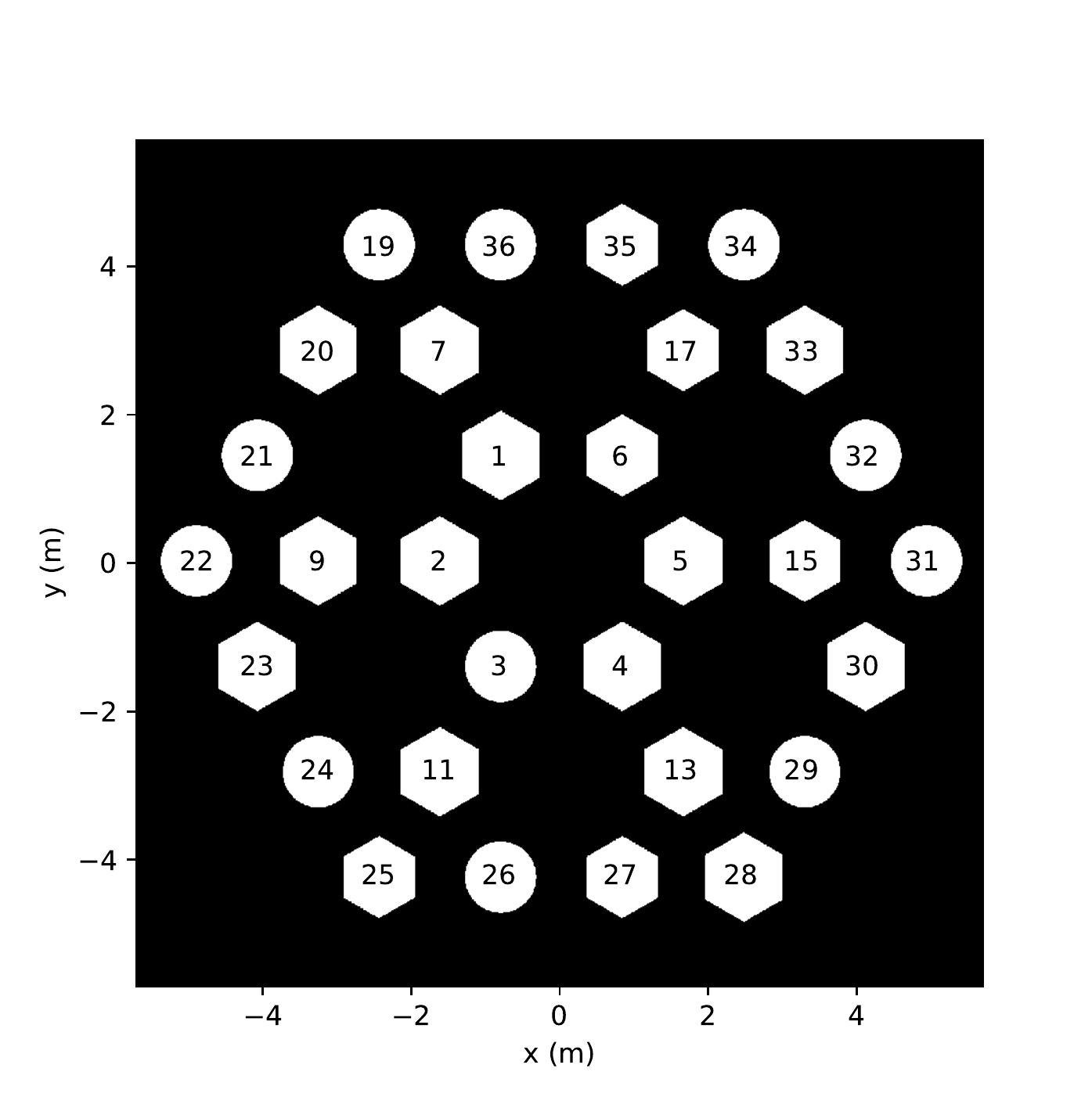}
        \caption{Numbering of the subapertures in the pupil mask.}
    \label{fig:HAM_pup_num}
\end{figure}

\begin{figure}
    \centering
    \includegraphics[trim=0 0 0 0,clip,width=0.45\textwidth]{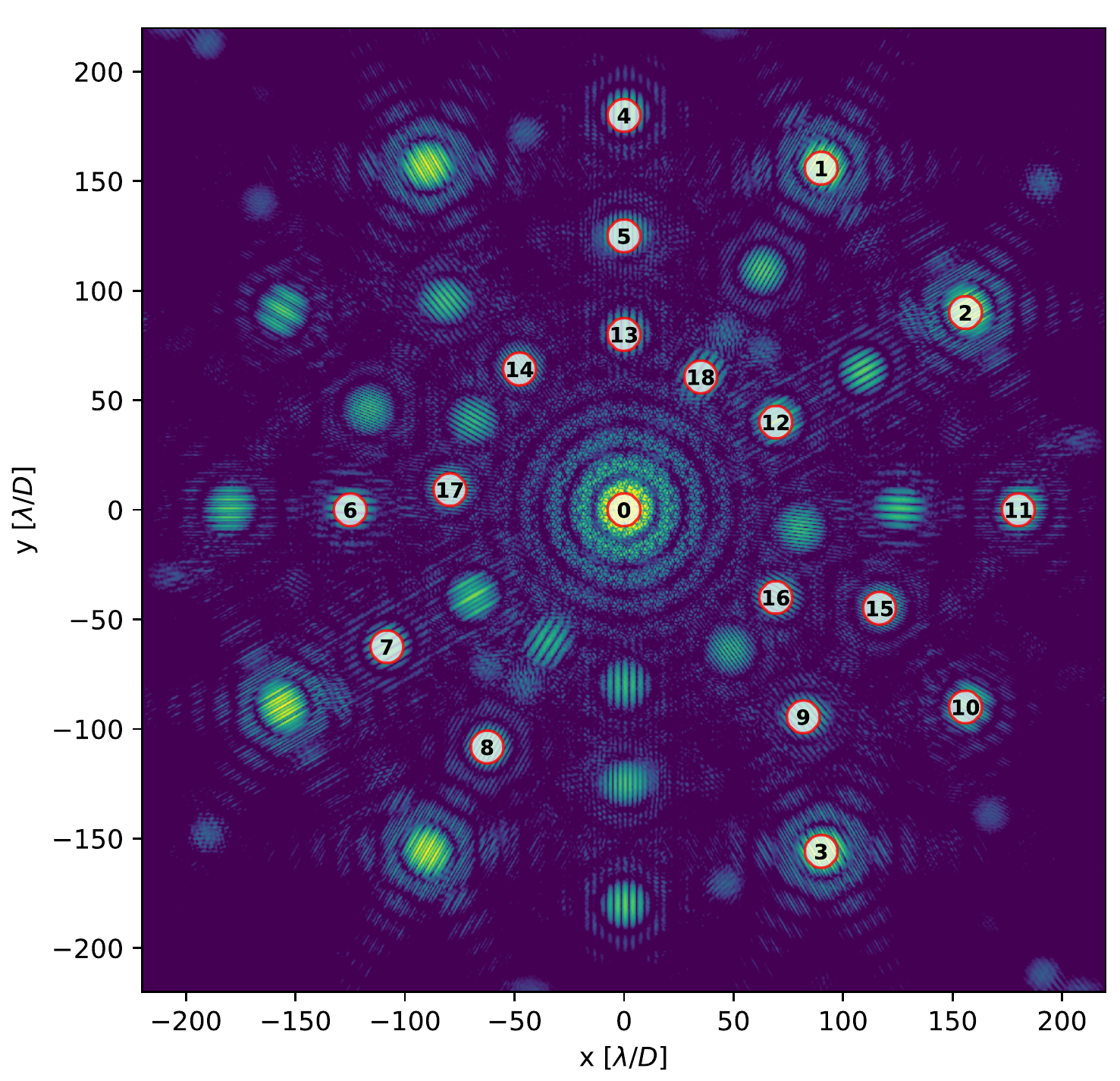}
        \caption{Numbering of the holographic PSFs.}
    \label{fig:HAM_PSF_num}
\end{figure}

Here we detail the full design parameters of the HAM mask for the OSIRIS instrument. The numbering of the subapertures is given in Fig. \ref{fig:HAM_pup_num}, and the numbering of the PSFs is given in Fig. \ref{fig:HAM_PSF_num}.
In Table \ref{tab:pupil_mapping} we give an overview of the subapertures and how they map to the PSFs or interferograms in the image plane. 
The PSF locations are given in Table \ref{tab:focal_plane_mapping}. 
The five different subaperture combinations of the holographic component are given in Fig. \ref{fig:HAM_subaper}. 
\begin{table}[h]
        \centering 
        \captionsetup{justification=centering}
        \caption{Subaperture locations and the numbers of the interferogram they  are imaged onto. }
        \vspace{-5pt}
        \begin{threeparttable}
                \begin{tabular}{l c c c c c}
                \hline \hline
            Comp& Num   &D (m)  &x (m)& y (m)&PSF num\\ 
            \hline

            SAM & & &&\\
            &   3       &0,939  &0,779  &1,350&0\\
            &   19      &0,939  &2,338  &-4,050&0\\
            & 21        &0,939  &3,897  &-1,350&0\\
            &   22      &0,939  &4,677  &0,000&0\\
            &   24      &0,939  &3,118  &2,700&0\\
            & 26        &0,939  &0,779  &4,050&0\\
            &   29      &0,939  &-3,118 &2,700&0\\
            &   31      &0,939  &-4,677 &0,000&0\\
            &   32      &0,939  &-3,897 &-1,350&0\\
            &   34      &0,939  &-2,338 &-4,050&0\\
            &   36      &0,939  &0,779  &-4,050&0\\
            HAM & & &&\\         
            A   &1      &1,250  &0,779  &-1,350&1\\
            A   &20     &1,250  &3,118  &-2,700&1\\
            A   &30     &1,250  &-3,897 &1,350&1\\
            B   &4      &1,250  &-0,779 &1,350&2\\
            B   &7      &1,250  &1,559  &-2,700&2\\
            B   &28     &1,250  &-2,338 &4,050&2\\
            C   &2      &1,250  &1,559  &0,000&3\\
            C   &23     &1,250  &3,897  &1,350&3\\
            C   &33     &1,250  &-3,118 &-2,700&3\\
            D   &6      &1,084  &-0,779 &-1,350&10,11,18\\
            D   &15     &1,084  &-3,118 &0,000&12,14,16,18\\
            D   &17     &1,084  &-1,559 &-2,700&10,12,17\\
            D   &25     &1,084  &2,338  &4,050&10,13,14,15\\
            D   &27     &1,084  &-0,779 &4,050&11,13,16,17\\
            D   &35     &1,084  &-0,779 &-4,050&11,12,15\\
            E   &5      &1,250  &-1,559 &0,000&5,6,9\\
            E   &9      &1,250  &3,118  &0,000&5,7,8\\
            E   &11     &1,250  &1,559  &2,700&4,7,9\\
            E   &13     &1,250  &-1,559 &2,700&4,6,8\\
        \hline
                \end{tabular}
        \end{threeparttable}
        \label{tab:pupil_mapping}
\end{table}

\begin{table}[h]
        \centering 
        \captionsetup{justification=centering}
        \caption{PSF numbers, their focal plane locations, and the subaperture numbers that are mapped there. The radii are provided in $\lambda/D$ as their locations change with wavelength due to diffraction.}
        \vspace{-5pt}
        \begin{threeparttable}
                \begin{tabular}{c c c c}
                \hline \hline
        PSF num & r ($\lambda/D$) & $\theta$ ($^\circ$)&Num\\
        \hline
        0&0&0&3,19,21,22,24,26 \\
        &&&29,31,32,34,36\\
        1& 180 & 60& 1,20,30\\
        2& 180 & 30& 4,7,28\\
        3& 180 & -60& 2,23,33\\
        4& 180& 90 & 11,13 \\
        5& 125& 90 & 5,9 \\
        6& -125 & 0 & 5,13 \\
        7& -125 & 30 & 9,11 \\ 
        8& -125 & 60 & 9,13 \\
        9& -125 & 130.89 & 5,11\\
        10& 180& -39 & 6,17,25\\
        11& 180 &0 & 6,27,35\\
        12& 80 & 30 & 15,17,35\\
        13 & 80 & 90 & 25,27\\
        14& 80 & 126.59 & 15,25\\
        15&-125&158.95& 25,35\\
        16&80&-30 & 15,27 \\
        17&80&173.41&17,27\\
        18&70&60&6,15\\
        \hline
                \end{tabular}
        \end{threeparttable}
        \label{tab:focal_plane_mapping}
\end{table}

\section{Impact of polarization leakage on the HAM OSIRIS design}
\label{app:leakage}

\rch{This appendix is a continuation of the analysis of the impact of polarization leakage on HAM, as described in Sect. \ref{sec:leakage_influence}. 
Polarization leakage emerges when the retardance of a geometric phase hologram is not exactly half-wave. 
As mentioned before, the geometric phase hologram is a half-wave retarder with a spatially varying fast axis. 
The space-variant Jones matrix of such a retarder in the circular polarization basis is described in \citet{ruane2019scalar} and \citet{doelman2020minimizing}, and it is given by
\begin{equation}
    \rch{M} = c_V \begin{bmatrix} 0 & \text{e}^{i2\chi(x,y)} \\ \text{e}^{-i2\chi(x,y)} & 0 \text{ } \end{bmatrix} + c_L \begin{bmatrix} 1 \text{ } & 0 \\ 0 \text{ } & 1 \end{bmatrix}.
    \label{eq:GPH}
\end{equation}
Here, $\chi(x,y)$ is the spatially varying fast-axis orientation; both $c_V$ and $c_L$ are parameters that depend on the retardance, $\Delta \phi$ \citep{mawet2009optical,ruane2019scalar}, and are given by
\begin{equation}
c_V = \sin{\frac{\Delta \phi}{2}},\text{  } c_L = -i \cos{\frac{\Delta \phi}{2}}.
\end{equation}
The first term in Eq. \ref{eq:GPH} describes that a fraction, $C_V$, of the light acquires a geometric phase of $\Phi(x,y) = \pm 2\chi(x,y)$,
where the sign of the phase depends on the handedness of the incoming circular polarization. 
The second term describes the polarization leakage beam, and it is unaffected by the fast-axis orientation pattern.
With \rch{imperfect} retardance, the output electric field for incoming right-circular polarization is given by
\begin{equation}
\mathbf{LC_{out}} = \rch{M} \mathbf{RC_{in}} = \rch{M} \begin{bmatrix} 1 \\ 0 \end{bmatrix} =  \begin{bmatrix} c_L \\ c_V\text{e}^{-i\Phi(x,y)} \end{bmatrix}.
\label{eq:leakage}
\end{equation}
The leakage term and main beam have an orthogonal polarization state and are therefore incoherent, assuming no polarization cross-talk caused by the optical system. 
In addition, unpolarized light contains, on average, equal amounts of left- and right-circular polarization, and these states are incoherent \citep[see][]{hecht1974optics}.
Therefore, we can describe the impact of leakage, typically on the order of 1\%, for one circular polarization state without loss of generality.
While objects do not appear fully unpolarized due to instrumental polarization or interstellar polarization, we first explore this simplification because it demonstrates how closure phases and complex visibilities are less resistant to wavefront aberrations.

From Eq. \ref{eq:leakage}, it is clear that the PSF is actually an incoherent sum of the HAM PSF and the leakage PSF, where the leakage PSF is the unaltered PSF from the HAM amplitude mask. 
Moreover, the Fourier transform is a linear operator. 
Therefore, we can calculate the visibilities of the HAM PSF and the leakage PSF separately and co-add them.
The combined measured visibility is then given by
\begin{equation}
V(\mathbf{f}) = c_V^2 V_{SAM}(\mathbf{f}) + c_L^2 V_{holo}(\mathbf{f}).
\end{equation}
Here, $V_{SAM}(\mathbf{f})$ is different from  $V_{holo}(\mathbf{f})$ since their amplitude masks are different.
For a broadband liquid-crystal phase mask, $c_V^2/c_L^2 < 3\%$ over a large bandwidth. 
However, for the HAM/OSIRIS design, the number of subapertures that are imaged off-axis is large (i.e., 19 versus the 11 of SAM). 
Combined, the 30 apertures contribute to many baselines multiple times.
The vector sum is entirely different for different aberrations and might even add up to zero for some baselines.

It is not correct to assume that the leakage PSF is incoherent for any linearly polarized light fraction.
Linearly polarized light can also be written as the sum of the two circular polarization states; however, they are still coherent.
This means that the polarization leakage of left-circular polarization interferes with the main beam of the right-circular polarization state that becomes left-circularly polarized after going through the half-wave retarder, that is,\begin{equation}
\mathbf{RC_{out}} = \rch{M} \mathbf{LP_{in}} =\frac{1}{\sqrt{2}} \rch{M} \begin{bmatrix} 1 \\ -i \end{bmatrix} =  \begin{bmatrix} c_L - i c_V\text{e}^{i\Phi(x,y)} \\ c_V\text{e}^{-i\Phi(x,y)} - i c_L\end{bmatrix}.
\end{equation}
The PSF will contain an interference term with a relative intensity of $\sim C_VC_L$, which is smaller than $C_V^2$ -- but much larger than $C_L^2$ -- when the retardance is close to half-wave } 

\begin{figure}[h!]
    \centering
    \includegraphics[width=\linewidth]{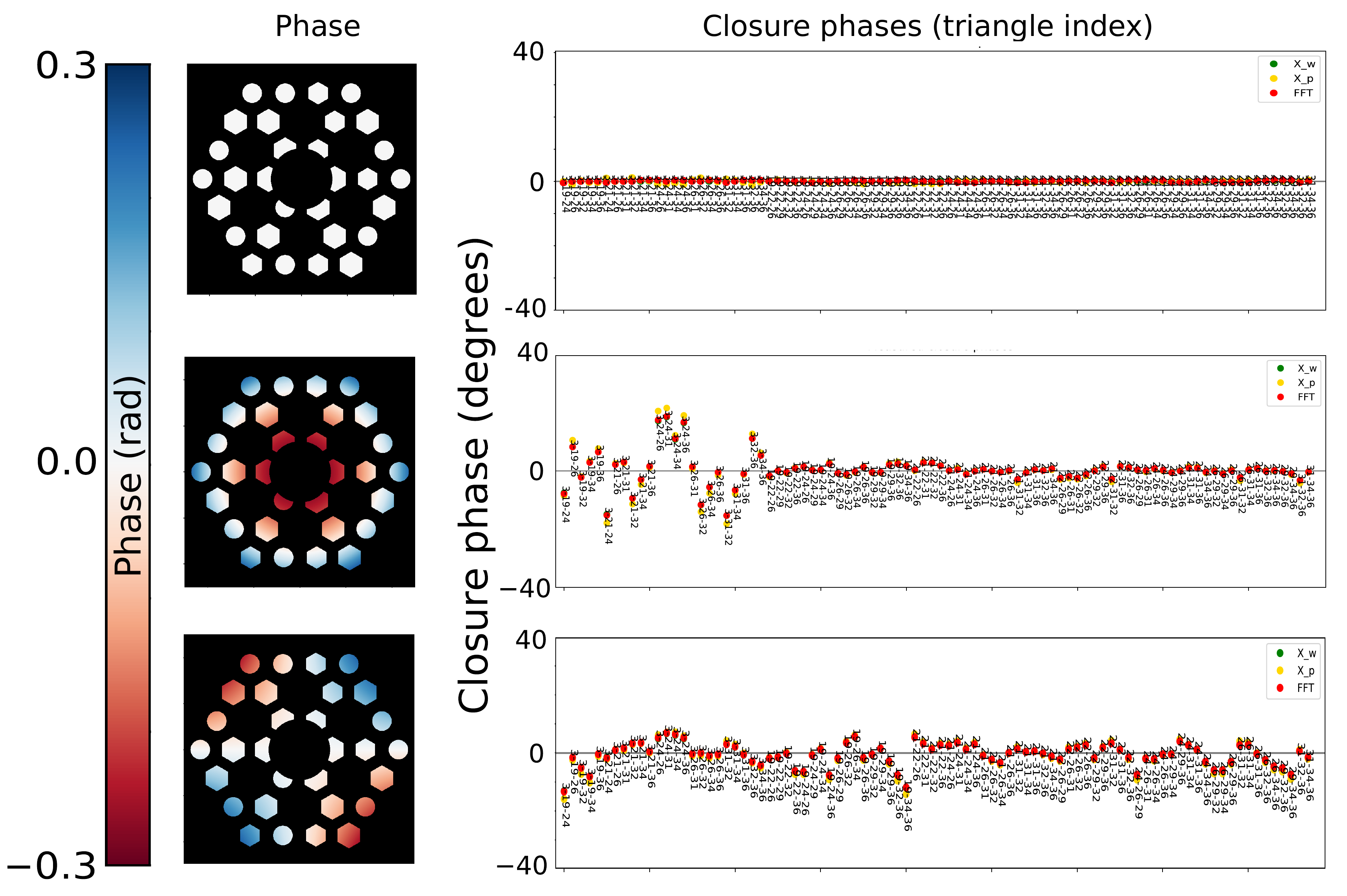}

        \caption{Closure phases of the central component for different wavefront aberrations, assuming 1\% polarization leakage.}
        \label{fig:leakage_wavefront_unpol}
\end{figure}

Next, we explore the impact of polarization leakage on the HAM OSIRIS design with simulations using the newly implemented polarization module of \texttt{HCIPy}.
The HAM optic is implemented as a spatially variable Jones matrix of a retarder with varying fast-axis orientation.
The electric fields are described by spatially variable Jones vectors. 
We varied the input polarization state and the phase aberrations of the wavefront before going through the HAM optic and the retardance of the HAM mask.
Total intensity PSFs are calculated with a plate scale similar to the laboratory results presented in this paper and without any detector noise. 
These PSFs are used as input for the HAM data reduction pipeline that calculates the closure phases of the central component only.
We show the dependence of the closure phases on wavefront aberration by simulating an un-aberrated wavefront and two aberrated wavefronts with 0.3 radian peak-to-valley defocus and astigmatism, respectively. 
The leakage is assumed to be 1\%, and the input polarization is unpolarized.
The results are shown in Fig. \ref{fig:leakage_wavefront_unpol}.
The simulations clearly indicate a strong dependence on wavefront aberrations. 
This effect is linear with polarization leakage and the strength of the aberration.
How each baseline is affected is also different for each aberration, and most of them average out to within a few degrees.

\begin{figure}[h!]
    \centering
    \includegraphics[width=0.93\linewidth]{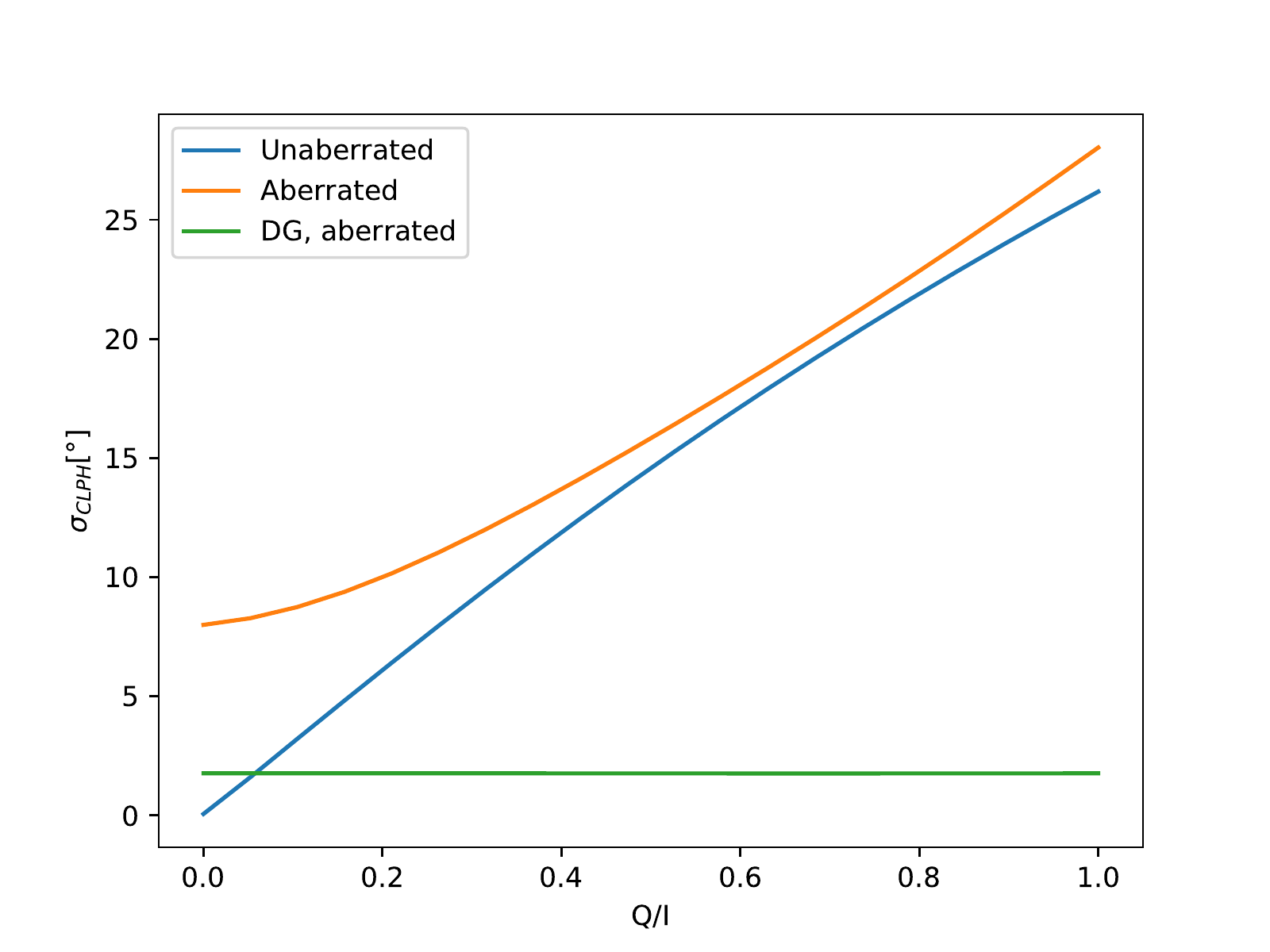}

        \caption{Scatter in the closure phases of the central component as a function of the linear polarization fraction. The wavefront is aberrated with defocus and astigmatism for an RMS of 48 nm at $\lambda = 1600$ nm. }
        \label{fig:leakage_lin_pol}
\end{figure}

Next we explore the impact of polarization leakage for linearly polarized light. 
As mentioned in Sect. \ref{sec:leakage_influence}, linearly polarized light has an even bigger impact on the closure phases.
The phase-shifted holes of the central component PSF and the zero-phase polarization leakage are coherent, causing a cross-term between the components that has a larger intensity than the incoherent leakage term.
This cross-term is independent of the incoherent leakage term, and, as such, the dependence on wavefront aberration as described for unpolarized light is still present for linearly polarized light.
However, the extra term does introduce closure phase offsets, even when the incoming wavefront is not aberrated.
The interference between many zero-phase leakage baselines with the phase-shifted central component baseline creates a nonzero closure phase that depends on the degree of linear polarization and the retardance of the HAM optic.
We simulated the impact of linearly polarized light with 2.5\% leakage, with and without aberrations, changing the fraction of linearly polarized light, Q.
The results are shown in Fig. \ref{fig:leakage_lin_pol}.
For small polarization fractions (<25\%), the impact is smaller or of similar magnitude compared to the unpolarized case with wavefront aberrations. 
Large linear polarization fractions induce large variations in the closure phases with the current HAM versions.
For astronomical objects with significant polarization fractions, the results would complicate the extraction of astrophysical parameters. 
It is difficult to calibrate these offsets since unresolved calibrators have a nonzero polarization fraction due to interstellar dust.
In Sect. \ref{sec:lab_tests} we estimated the degree of linear polarization and fit low-order phase aberrations to the measured closure phases using a forward model.
However, we did not measure these values independently to see if the recovered values agreed with the aberrations present in the system.
A true determination of the offsets caused by linearly polarized light from the source, instrumental polarization, or wavefront aberrations would require a full Mueller matrix model of the instrument as well as focal plane wavefront sensing independent of the HAM PSF. 

Reducing the polarization leakage, either by filtering or by using the double-grating method, would remove the need for any postprocessing solution.
The simulation using a double-grating HAM in Fig. \ref{fig:leakage_lin_pol} shows that $\sigma_{CLPH}$ is almost completely independent of the linear polarization fraction. 
However, it is nonzero due to wavefront aberrations. 
This could point at minor inaccuracies in the closure phase reconstruction.
The scatter caused by the coherent leakage term and the central component is much larger in simulations than the theory would predict with the extremely small leakage fraction of $6.25 \times 10^{-4}$.
Overall, it is clear that the double-grating method would dramatically improve the performance of the central component of HAM.
We aim to incorporate the double-grating method in a future upgrade of HAM for the OSIRIS Imager.


\end{document}